%% file: 0paper.tex
\documentclass[acmlarge, screen, nonacm]{acmart}

\input{00packages.tex}

\AtBeginDocument{%
  \providecommand\BibTeX{{%
    \normalfont B\kern-0.5em{\scshape i\kern-0.25em b}\kern-0.8em\TeX}}}

\setcopyright{none}
\begin{document}

\title[Interpersonal Theory of Suicide as a Lens to Examine Suicidal Ideation in Online Spaces]{Interpersonal Theory of Suicide as a Lens to Examine Suicidal Ideation in Online Spaces}


\author{Soorya Ram Shimgekar}
\orcid{0000-0002-1110-9699}
\affiliation{%
 \institution{University of Illinois Urbana-Champaign}
 \city{Urbana}
 \state{IL}
 \country{USA}}
 \email{sooryas2@illinois.edu}

 \author{Violeta J. Rodriguez}
\orcid{0000-0001-8543-2061}
\affiliation{%
 \institution{University of Illinois Urbana-Champaign}
 \city{Urbana}
 \state{IL}
 \country{USA}}
 \email{vjrodrig@illinois.edu}

 \author{Paul A. Bloom}
\orcid{0000-0003-3970-5721}
\affiliation{%
 \institution{Columbia University Irving Medical Center, New York State Psychiatric Institute}
 \city{New York}
 \state{New York}
 \country{USA}}
 \email{paul.bloom@nyspi.columbia.edu}

 \author{Dong Whi Yoo}
\orcid{0000-0003-2738-1096}
\affiliation{%
 \institution{Kent State University}
 \city{Kent}
 \state{OH}
 \country{USA}}
 \email{dyoo@kent.edu}

\author{Koustuv Saha}
\orcid{0000-0002-8872-2934}
\affiliation{%
 \institution{University of Illinois Urbana-Champaign}
 \city{Urbana}
 \state{IL}
 \country{USA}}
 \email{ksaha2@illinois.edu}

\renewcommand{\shortauthors}{Shimgekar et al.}


\begin{abstract}

Suicide is a critical global public health issue, with millions experiencing suicidal ideation (\sci{}) each year. 
Online spaces enable individuals to express \sci{} and seek peer support. 
While prior research has revealed the potential of detecting \sci{} using machine learning and natural language analysis, a key limitation is the lack of a theoretical framework to understand the underlying factors affecting high-risk suicidal intent. To bridge this gap, we adopted the \textit{Interpersonal Theory of Suicide} (\its{}) as an analytic lens to analyze 59,607 posts from Reddit's \textit{r/SuicideWatch}, categorizing them into \sci{} dimensions (Loneliness, Lack of Reciprocal Love, Self Hate, and Liability) and risk factors (Thwarted Belongingness, Perceived Burdensomeness, and Acquired Capability of Suicide).
We found that high-risk \sci{} posts express planning and attempts, methods and tools, and weaknesses and pain.
In addition, we also examined the language of supportive responses through psycholinguistic and content analyses to find that individuals respond differently to different stages of Suicidal Ideation (\sci{}) posts.
Finally, we explored the role of AI chatbots in providing effective supportive responses to suicidal ideation posts.
We found that although AI improved structural coherence, expert evaluations highlight persistent shortcomings in providing dynamic, personalized, and deeply empathetic support. 
These findings underscore the need for careful reflection and deeper understanding in both the development and consideration of AI-driven interventions for effective mental health support.

\end{abstract}




\maketitle

\input{1introduction} 
\input{2related} 
\input{3data} 
\input{4Aim1} 
\input{5Aim2} 
\input{6Aim3} 
\input{7Aim4} 
\input{8discussion}
\input{9conclusion} 


\bibliographystyle{ACM-Reference-Format}
\bibliography{0paper}




\end{document}
\endinput

%% file: 00packages.tex
\usepackage{color, colortbl, xcolor}
\usepackage{url}
\usepackage{subcaption}
\usepackage{textcomp}
\usepackage{soul}
\usepackage{multirow}
\usepackage{enumitem}
\usepackage{mathtools}
\usepackage{siunitx}

\usepackage{graphicx} 
\usepackage{placeins} 

\usepackage{subcaption}
\usepackage{graphicx}
\usepackage{lipsum}
\usepackage{caption}

\usepackage{array} 
\usepackage{ragged2e} 

\usepackage{tikz}
\usepackage{xcolor}

\usepackage{booktabs} 
\usepackage{graphicx} 
\usepackage{graphicx}  

\usepackage{arydshln}
\definecolor{linkColor}{RGB}{6,125,233}
\definecolor{green}{rgb}{0.0, 0.65, 0.31}
\definecolor{bleudefrance}{rgb}{0.19, 0.55, 0.91}
\definecolor{ceruleanblue}{rgb}{0.16, 0.32, 0.75}
\definecolor{grey}{HTML}{969696}
\definecolor{violet}{HTML}{756bb1}
\definecolor{dgrey}{HTML}{525252}
\definecolor{lgrey}{HTML}{5ab4ac}
\definecolor{dgreen}{HTML}{005a32}
\definecolor{purple}{HTML}{ae017e}

\definecolor{lethalcolor}{HTML}{e66101}
\definecolor{nlethalcolor}{HTML}{5e3c99}

\def\dgreybar#1{{\color{dgrey}\rule{#1pc}{6pt}}}

\definecolor{editCol}{HTML}{000000}
\definecolor{maskCol}{HTML}{c51b7d}
\definecolor{lrColor}{HTML}{8856a7}
\definecolor{trColor}{HTML}{d01c8b}
\definecolor{ctColor}{HTML}{4dac26}
\definecolor{brickred}{HTML}{f03b20}
\definecolor{improveCol}{HTML}{253494}
\definecolor{worsenCol}{HTML}{d7191c}
\definecolor{DarkBlue}{HTML}{00008B}
\definecolor{mscolor}{HTML}{01665e}
\definecolor{nmscolor}{HTML}{bf812d}
\definecolor{lgreen}{HTML}{ccece6}
\definecolor{dolive}{HTML}{308014}
\definecolor{lgreen}{HTML}{e0f3db}
\definecolor{dpink}{HTML}{CD1076}
\definecolor{pink}{HTML}{FED2D2}
\definecolor{soothinggreen}{HTML}{4dac26}
\definecolor{darkred}{HTML}{8B0000}

\colorlet{tablerowcolor4}{gray!50} 

\newcommand*{\textlabel}[2]{%
  \edef\@currentlabel{#1}
  \phantomsection
  #1\label{#2}
}

\colorlet{tableheadcolor}{gray!25} 
\colorlet{tablerowcolor}{gray!10} 
\colorlet{tablerowcolor2}{gray!45} 
\colorlet{tablerowcolor3}{gray!25} 

\newcommand{\rowcollight}{\rowcolor{tablerowcolor3}} %

\newcolumntype{a}{>{\columncolor{tablerowcolor}}r}
\definecolor{aicolor}{HTML}{018571}
\definecolor{occolor}{HTML}{ff7799}

\definecolor{aicolor}{HTML}{fc8d62}
\definecolor{occolor}{HTML}{253494}

\newif{\ifhidecomments}
  \hidecommentsfalse 
\ifhidecomments
    \newcommand{\soorya}[1]{}
    \newcommand{\violeta}[1]{}
    \newcommand{\paul}[1]{}
    \newcommand{\dongwhi}[1]{}
    \newcommand{\koustuv}[1]{}
\else
    \newcommand{\soorya}[1]{\textbf{\small\sffamily{\textcolor{DarkBlue}{[#1 -- Soorya]}}}}
    \newcommand{\violeta}[1]{\textbf{\small\sffamily{\textcolor{orange}{[#1 -- Violeta]}}}}
    \newcommand{\paul}[1]{\textbf{\small\sffamily{\textcolor{dolive}{[#1 -- Paul]}}}}
    \newcommand{\dongwhi}[1]{\textbf{\small\sffamily{\textcolor{marroon}{[#1 -- Dong Whi]}}}}
    \newcommand{\koustuv}[1]{\textbf{\small\sffamily{\textcolor{dpink}{[#1 -- Koustuv]}}}}
  \fi

\newcommand{\sci}{\textsf{SI}}
\newcommand{\its}{\textsf{IPTS}}
\newcommand{\swatch}{\textit{r/SuicideWatch}}
\newcommand{\sdim}{\textsf{Dimension}}
\newcommand{\srsk}{\textsf{RiskFactor}}

\newcommand{\ai}{AI}
\newcommand{\oc}{OC}

\renewcommand{\textrightarrow}{$\rightarrow$}

\colorlet{tableheadcolor}{gray!25} 

\definecolor{neutralCol}{HTML}{dd1c77}
\definecolor{neutralGreen}{HTML}{31a354}
\definecolor{NewBlue}{HTML}{1879ba}
\definecolor{bleudefrance}{rgb}{0.19, 0.55, 0.91}  
\definecolor{AfTrColor}{HTML}{0868ac}  
\definecolor{BfTrColor}{HTML}{a8ddb5}  

\definecolor{AfCtColor}{HTML}{b10026}  
\definecolor{BfCtColor}{HTML}{fd8d3c}

\graphicspath{ {figures/} }

\newcommand{\para}[1]{\vspace{0.3em}\noindent\textbf{#1}~}

%% file: 1introduction.tex
\section{Introduction}

Amid the escalating mental health concerns worldwide, suicide has emerged as a major public health crisis, claiming approximately 700K deaths each year, with a disproportionate impact on young adults and marginalized communities~\cite{who_suicide}. Beyond fatalities, millions more experience suicidal ideation (\sci{}) or attempts, further exacerbating the mental health burden worldwide~\cite{harmer2020suicidal}.
In this context, it is critical to find safe spaces for individuals to express \sci{} and receive timely and effective support and intervention. 
The widespread adoption of the Internet and digital tools
has facilitated the prevalence of dedicated online spaces where individuals can share mental health struggles and seek peer support.
Such online support tools can be based on human-human interactions, such as instant messaging and social media platforms~\cite{thurlow2004computer, meier2021computer}, or human-AI interactions, including chatbots~\cite{gudala2022benefits,ayers2023comparing,xu2021chatbot}.
Given socioeconomic disparities, limited access to mental health services, and pervasive stigma, online support tools are especially advantageous in offering several key benefits, including anonymity, peer support, and the flexibility of asynchronous participation~\cite{de2014mental}.
Prior work noted how these online communities can foster
social support, empathy, and connection~\cite{de2014mental, dutta2021temporal,andalibi2018social,grattidge2023exploring, wadden2021effect}.
Additionally, we are seeing the rise of generative AI-driven chatbots, which can provide immediate, AI-driven, human-like responses to mental health queries~\cite{song2024typing, chen2023llm,chang2025effectiveness,escobar2023chatbot}.
Despite these developments, a critical gap remains in understanding the theoretical foundations of online interactions in high-stakes contexts, particularly \sci{}.

Prior work has highlighted the effectiveness of online communities, such as on Reddit, TalkLife, and 7cups, in providing spaces where individuals can discuss, seek, and share information, advice, and social support related to mental health concerns~\cite{saha2020omhc,de2014mental,olteanu2017distilling,yang2019channel,smith2020cannot}.
These platforms, with features of anonymity (or pseudonymity), moderation, and structured peer-support interactions, foster safe spaces that encourage candid and sensitive self-disclosures and promote a sense of belonging and solidarity among peer supporters. 
Earlier research highlights that participation in moderated peer-support spaces can improve mental well-being~\cite{wadden2021effect, sharma2018mental,saha2020causal}.
Relatedly, De Choudhury and colleagues examined 
the language of social support in response to \sci{} in online communities~\cite{de2017language}.

Despite advancements in detecting \sci{} in online spaces~\cite{yeskuatov2024detecting, ji2018supervised, liu2022detecting, alghazzawi2025explainable}, a critical gap remains in applying a theoretical lens to understand the mechanisms underlying these interactions. 
Given that suicidal thoughts rarely emerge in isolation, a more nuanced approach is needed---one that accounts for the psychological and social factors influencing suicidal progression~\cite{szanto2002identification}.
The Interpersonal Theory of Suicide (\its{}) provides a well-tested framework for understanding the mechanisms of both \sci{} and transitions from ideation to suicidal behaviors (e.g. ``ideation-to-action''~\cite{klonsky2018ideation}), shedding light into the different \sdim{}s and \srsk{}s of \sci{}~\cite{van2010interpersonal, joiner2010interpersonal}.
\its{} posits that suicidal behavior is most likely when three key psychological \srsk{}s converge: Thwarted Belongingness, Perceived Burdensomeness, and Acquired Capability for Suicide~\cite{joiner2010interpersonal}.
Given the growing reliance on online communities for mental health support, applying this theory to online discourse could enhance our understanding of suicide risk assessment and intervention strategies in digital settings.

It is important to offer immediate, around-the-clock,  portable assistance to individuals suffering from \sci{}. A plausible solution to this could be AI chatbots, which provide spaces for personal and interactive journaling, as well as educational resources for self-help and coping strategies~\cite{haque2023overview,van2023providing,balcombe2023ai,thakkar2024artificial}.
Although such tools hold promise for supplementing traditional therapy, concerns have been raised regarding their effectiveness and the importance of maintaining human oversight in mental health care~\cite{cho2023integrative}.
Furthermore, effective response strategies for different types of \sci{}, in terms of linguistic characteristics, remain underexplored.
Such an understanding can help improve timely and tailored online mental health interventions.

With the above motivation, this paper has the following research aims:

\begin{enumerate}
    \item \textbf{Aim 1}: Examine how \sci{} manifests in online self-disclosures through the lens of the \its{}.
        \item \textbf{Aim 2}: Analyze what linguistic cues are associated with responses to \sci{} disclosures in online spaces. 
    \item \textbf{Aim 3}: Evaluate the language of an AI chatbot's responses to online \sci{} disclosures.
\end{enumerate}



We conducted our study on 59,607 posts and 149,144 comments collected from the \swatch{} subreddit on Reddit---an online community dedicated to \sci{}-related discussions with over 516K members (as of February 2025). 
First, for Aim 1, we adopted a theory-driven lens of operationalizing \its{} within our dataset by employing unsupervised machine learning and iterative codebook development.
This approach allowed us to label the $\sim$59K posts based on \sdim{}s (loneliness, lack of reciprocal love, self-hate, and liability), as well as \srsk{}s (thwarted belongingness, perceived burdensomeness, and acquired capability), using a data-driven process informed by iterative codebook development.
Our computational approach finally identified 1,508 lethally suicidal posts---those exhibiting all three risk factors together. 

Next, for Aim 2, we analyzed responses to \sci{} posts through psycholinguistic analyses (using Linguistic Inquiry and Word Count~\cite{pennebaker2001linguistic}) and content analyses (using Sparse Additive Generative Model~\cite{eisenstein2011sparse}), identifying key linguistic characteristics of responses to different kinds of \sci{} posts. 
Our analyses revealed that responses to thwarted belongingness contained more negative affect and tentative language. In contrast, responses to perceived burdensomeness were more positive but hesitant, and responses to acquired capability emphasized temporal references to build personal connections.





Finally, for Aim 3, we explored various prompting strategies for AI chatbots (using GPT-4o), incorporating \srsk{}s and key characteristics of supportive responses.
We compared the AI-generated responses to human-written responses in online communities, finding that although AI-generated responses consisted of better linguistic structure and semantic alignment to the original post, these responses were less diverse, more complex, and more formal compared to human-written responses. 
We also expert-validated these AI responses with our  psychologist co-authors to identify persistent limitations of providing genuine empathy.


This paper makes the following key contributions---1) a theory-driven computational framework to label online disclosures of \sci{},
2) a linguistic analysis of supportive responses to various \sci{} posts,
and 3) a preliminary evaluation of AI online disclosures of \sci{} posts through quantitative analysis and expert evaluation. 
This study underscores the value of incorporating a theory-based lens in building digital mental health interventions, demonstrating how established psychological frameworks can enhance suicide risk assessment and support strategies. 
We discuss the implications in responsible design and deployment of mental health interventions.
For instance, online platforms can integrate \its{}-based models to assess and triage critical cases of suicidal risks for timely intervention. 
Although AI-driven mental health support holds potential, its effectiveness hinges on its adaptability to individual needs and the presence of human oversight. 
Rather than functioning autonomously, AI should serve as a complementary tool that enhances, rather than replaces, human-led crisis interventions~\cite{sharma2023human, palmer2024combining, babu2024artificial}. 

\para{Privacy, Ethics, and Reflexivity.}
This paper used publicly accessible social media discussions on Reddit and did not require direct interactions with individuals, thereby not requiring an ethics board approval. 
However, we are committed to the ethics of the research and followed practices to secure the privacy of individuals in our dataset. 
This paper only presents paraphrased quotes to reduce traceability yet provide context in readership. 
Our research team comprises researchers holding diverse gender, racial, and cultural backgrounds, including people of color and immigrants, and hold interdisciplinary research expertise.
Our research team comprises computer scientists with expertise in HCI, CSCW, and social computing and psychologists with expertise in clinical psychology, adolescent depression and suicide, and digital health interventions.
One of our psychologist coauthor specialize in suicide etiology, suicide prevention, and crisis intervention, and the other psychologist coauthor is a clinical psychologist with over 16 years of experience spanning adult and adolescent inpatient care and crisis suicide helplines.
To ensure validity and prevent misrepresentation, our findings were reviewed and corroborated by our psychologist coauthors. 
However, our work is not intended to replace the clinical evaluation of an individual undergoing suicidal thoughts and should not be taken out of context to conduct mental health assessments. 
We further discuss the ethical implications of our work in~\autoref{sec:ethical_implications}.


%% file: 2related.tex
\section{Background and Related Work}\label{section:rw}

\subsection{Suicidal Ideation (\sci{}): Definition and Theory} 
Suicidal Ideation (\sci{}), encompassing thoughts of ending one’s life---a pressing public health concern that has garnered significant attention from researchers and mental health professionals~\cite{nock2008suicide}. 
Understanding the underlying mechanisms behind suicidal thoughts and behaviors is essential for developing effective intervention strategies~\cite{klonsky2016suicide, jobes2019reflections} . 
Prior research has emphasized the need for targeted assessment tools that extend beyond a generalized mental health approach~\cite{klonsky2016suicide, jobes2019reflections}. 
A critical need exists to identify individuals experiencing suicidal thoughts and those at risk of progressing to suicidal attempts~\cite{klonsky2016suicide,franklin2017risk}.

A popular and well-validated psychological framework that offers a structured perspective on \sci{} thoughts and behaviors is the \textbf{Interpersonal Theory of Suicide} (\its{})~\cite{van2010interpersonal,joiner2010interpersonal}.
, which
\its{} posits that individuals who feel socially disconnected (thwarted belongingness) or perceive themselves as a burden to others (perceived burdensomeness) are at an elevated risk of developing \sci{}. 
However, for these desires to translate into a suicidal attempt, \its{} emphasizes the role of acquired capability---the learned ability to endure pain and suppress the fear of death. 
This framework particularly applies to individuals with a history of self-injury, chronic illness, or exposure to traumatic experiences, as these factors can contribute to increased pain tolerance and reduced fear of dying, making a suicide attempt more likely~\cite{van2010interpersonal}.

Expanding on these ideas,~\citet{klonsky2021three} proposed the Three-Step Theory (3ST)~\cite{klonsky2021three}, suggesting that suicidal thoughts emerge when an individual's psychological pain and hopelessness surpass their sense of connectedness. 
3ST posits that for ideation to escalate into an attempt, individuals must also develop the capability to act---often through repeated exposure to painful or fear-inducing experiences, such as self-harm or previous suicide attempts. 
These insights about an individual experiencing different stages of \sci{}s align with the findings of~\citet{franklin2017risk}, which emphasize the need for personalized support based on each person's mental state.

While these psychological models provide valuable insights into the mechanisms of \sci{} and behavior, they do not operate in isolation.
The Integrated Motivational-Volitional (IMV) Model ~\cite{o2018integrated} enhances the understanding of suicide risk by delineating it into three distinct phases. The pre-motivational phase accounts for background vulnerabilities and triggering life events. The motivational phase explores feelings of entrapment and defeat that give rise to suicidal thoughts. Finally, the volitional phase identifies factors, such as access to means, that increase the likelihood of transitioning from suicidal thoughts to actions.
By distinguishing between the processes that lead to suicidal thoughts and those that lead to actions, the IMV model enhances suicide risk assessment and informs prevention strategies tailored to different stages of ideation~\cite{o2018integrated}.

That said, prior research also recognized that the different theories on suicidal ideation share many similarities while highlighting distinct mechanisms~\cite{kirshenbaum2024adolescents}.
Most studies testing the above-mentioned ideation-to-action theories of suicide rely on interviews or assessments (e.g., self-report) to measure suicidal thoughts and behaviors, and, it is critical to find alternative and proactive means of assessing suicidal ideation. 
Various standardized instruments, such as the Columbia-Suicide Severity Rating Scale (C-SSRS)~\cite{posner2008columbia} and the Patient Health Questionnaire (PHQ-9)~\cite{manea2015diagnostic}, have been developed to measure the severity of suicidal thoughts and behaviors. 
Although these tools have demonstrated effectiveness in identifying individuals at risk for suicide in clinical settings, they often depend on self-reported data, which may be affected by stigma, privacy concerns, fear of judgment, sociocultural differences or desire to avoid involuntary hospitalization or other intervention, resulting in the underreporting of \sci{} in real-world data collection~\cite{lowry2024suicide, brown2020c, roaten2021universal, shin2025suicide}. 
Additionally, conducting such assessments on a large scale requires substantial time, training, and resources, making widespread implementation challenging and resource-intensive. 

Drawing motivation from the above, this study adopts natural language analyses to identify suicidal ideation in online self-disclosures. In particular, we use the \its{} framework given its applicability among young people~\cite{kirshenbaum2024adolescents}---a population that is especially prevalent in online spaces.

\subsection{Mental Health Self-Disclosure and Support on Social Media}

With the widespread use of the Internet and digital devices, online spaces have become key platforms for mental health self-disclosure~\cite{de2014mental}. Communities like Reddit enable structured discussions in specialized subreddits, fostering targeted support and a sense of belonging. A rich body of prior work suggests that moderated peer-support spaces can enhance wellbeing~\cite{wadden2021effect,saha2020causal,de2015social,andalibi2017sensitive}. These platforms enable individuals to discuss stigmatized and underexplored aspects of their experiences, fostering connections and reducing feelings of isolation~\cite{johnson2022s}. 
Social support, a well-established construct in psychology, has been extensively studied in both offline and online contexts~\cite{loane2013communication,cutrona1986social}. 
The Social Support Behavioral Code (SSBC) identifies five key types of social support—emotional, informational, esteem, network, and tangible support—of which emotional and informational support are the most prevalent in online health communities~\cite{suhr2004social, sharma2018mental, de2017language, braithwaite1999communication, buis2008emotional, eriksson2000informational, nakikj2017park,saha2020causal,kim2023supporters,smith2020cannot}.

Psycholinguistic research has underscored the importance of language in seeking and providing social support, highlighting how linguistic markers impact mental health outcomes~\cite{pennebaker2007expressive, pennebaker2001linguistic, chung2007psychological}. 
Research in psychotherapeutic settings has emphasized the significance of factors such as empathy, warmth, congruence, and therapeutic alliance in predicting care-seekers' outcomes~\cite{lambert2001research, labov1977therapeutic, norcross2018psychotherapy,althoff2016large}. 
Translating these principles to digital mental health interventions, studies have explored the potential of computational methods to infer psychosocial dynamics from social media language, offering insights into distress detection and support mechanisms~\cite{chancellor2016quantifying, de2013predicting, guntuku2017detecting, saha2021person,chancellor2020methods}. Prior work has established the construct validity of these measures, demonstrating that social media discussions can predict visits to emergency and inpatient hospital services~\cite{guntuku2020variability} and counseling centers~\cite{saha2022social}.

Technology-aided health interventions have gained traction, with research demonstrating the effectiveness of computer-mediated psychotherapy in achieving positive outcomes~\cite{cavanagh2018my}. 
Social media-based mental health interventions have shown promise, with studies analyzing the linguistic characteristics of supportive responses and their impact on user wellbeing~\cite{ernala2017linguistic, sharma2020computational, merolli2013health,saha2020causal}. 
\citeauthor{althoff2016large} examined counseling strategies such as adaptability, creativity, and perspective change, and \citeauthor{saha2020causal} identified key response factors---adaptability, immediacy, diversity, and emotionality---associated with positive wellbeing outcomes in online peer-support spaces~\cite{althoff2016large, saha2020causal}. Further,~\citeauthor{sharma2020computational} developed an empathy-detection model for mental health support~\cite{sharma2020computational}

Particularly relevant to the space of suicidal ideation on social media, prior work has developed machine learning and computational approaches on detecting suicidal ideation in social media language~\cite{saha2019social,burnap2015machine,rabani2023detecting,thieme2020machine,ramirez2020detection}.~\citeauthor{al2018absolute} used LIWC for psycholinguistic classification, while \citeauthor{lee2018assessment} applied neural networks to identify recurring suicidal topics, inspiring our use of TopicBERT for topic clustering. \citeauthor{walker2023linguistic} found linguistic differences in Instagram captions between teenage suicide decedents and living controls, and \citeauthor{cheng2017assessing} showed that high-suicide-probability posts contain more pronouns and fewer verbs. Beyond detection, \citeauthor{coppersmith2016exploratory} identified emotional patterns around suicide attempts, and \citeauthor{de2016discovering} applied computational methods to Reddit data, isolating language markers of \sci{} through statistical analysis.

Our study builds on and contributes to the above body of work by investigating \sci{} disclosure and interactions in online spaces. 
We draw on prior work that adopted a theory-driven computational approach on social media data to understand social phenomenons~\cite{zhou2022veteran,razi2020let,saha2019language}. This study unveils the progression from suicidal thoughts to attempts, analyzing both distress signals and intervention need in online spaces. We also examine the linguistic variations in supportive responses across different \sci{} types, offering insights into crisis interventions.

\subsection{AI Chatbots for Mental Health}

In an era where mental health crises are on the rise, accessible support systems have never been more critical. Traditional therapy remains the gold standard, yet the growing demand for mental health resources far exceeds the supply of professionals available to provide care. In response, online support tools including AI-driven chatbots, have emerged as promising alternatives. These technologies offer immediate, around-the-clock support, supplying users with coping mechanisms and empathetic engagement at the touch of a button~\cite{haque2023overview,van2023providing,balcombe2023ai,thakkar2024artificial}. While such advancements hold potential as supplementary interventions, ethical concerns and the necessity for human oversight remain key points of discussion~\cite{cho2023integrative}.  

AI chatbots have showed promise in addressing depression and distress, but challenges in usability, engagement, and healthcare integration persist~\cite{abd2020effectiveness}. Personalization and context-specific adaptation are key to improving efficacy~\cite{casu2024ai, li2023systematic}. Prior work has used attention-based transformers and sentiment analysis to provide tailored, anonymous mental health support~\cite{bhagchandani2022deep}.
For example, research on the AI companion ``Replika,'' which utilizes GPT-3 and GPT-4, suggests that such models significantly improve user well-being and play a role in suicide prevention~\cite{maples2024loneliness}. Despite these promising results, large-scale clinical trials remain necessary to evaluate their sustained efficacy and potential integration into traditional mental healthcare~\cite{omarov2023artificial,zhang2024dark}.  

Recent research has sought to understand the user experience of interacting with chatbots in the context of mental health support~\cite{sharma2024facilitating,sweeney2021can,vaidyam2019chatbots}. While some individuals report positive interactions, risks persist, particularly when these chatbots are not meticulously designed~\cite{song2024typing}. Studies indicate that LLM-based chatbots can exhibit empathetic behaviors and ask thorough questions about mental health symptoms, such as sleep disturbances. However, they fall short in ruling out associated conditions compared to human psychiatrists, raising concerns about potential misdiagnosis and the overall depth of AI-driven interventions~\cite{lawrence2024opportunities}. 

Despite technological advancements, skepticism persists within the mental health profession. Concerns over reliability, ethical considerations, and liability contribute to hesitancy in fully embracing AI-powered interventions~\cite{stapleton2024if}. 
Despite promising advancements, LLMs in mental health care raise ethical concerns, including accuracy, emotional distress, and over-reliance on technology ~\cite{de2023benefits,cabrera2023ethical}.
Addressing these challenges through rigorous validation and collaboration with both mental health experts and policymakers will help ensure responsible AI integration in therapeutic contexts. To improve AI responses to \sci{} disclosures, it is crucial to identify the most effective prompts and understand how LLMs could address \sci{} \srsk{} and \sdim{}. 
Our study examines the language of AI chatbots to \sci{} disclosures.
We validate our findings with experts to clinically assess AI responses across different prompt styles and identify potential risks. 
By evaluating the strengths and limitations of LLM-driven interventions, our work provides insights to enhance AI support systems and bridge the gap between detection and intervention.

%% file: 3data.tex
\section{Data}\label{section:data}

We sourced our data on \sci{} from the subreddit \swatch{} on Reddit. 
Reddit is a widely used semi-anonymous social platform consisting of online communities, called subreddits, which are dedicated to specific themes of discussions and topics.
Prior work has obtained and studied Reddit data for suicidal ideation~\cite{de2016discovering,de2017language} as well as other mental health concerns~\cite{sharma2018mental,saha2020omhc,de2014mental}. 
This body of research established that design features such as pseudonymity, community-driven moderation, and asynchronous peer support on Reddit enable individuals to overcome mental health related stigma and candidly self-disclose their sensitive mental health concerns and seek social support from other community members~\cite{de2014mental,saha2020omhc,andalibi2018social,yan2024identifying}.
Essentially, Reddit has several communities dedicated to mental health discussions~\cite{sharma2018mental,saha2020omhc}, and among these, the subreddit \swatch{} self-describes itself as, ``peer support for anyone struggling with suicidal thoughts.'' 
The subreddit was started on December 16, 2008, and consists of over 516,000 members as of February 2025. 
This subreddit is heavily moderated with eight active moderators, and the community guidelines explicitly prohibit harmful responses, tough love, guilt-tripping, and actions like trolling or promoting suicide. Further, the community advises against recommending specific therapies, self-help strategies, or medications.

We collected the Reddit data using a publicly available Reddit API from \textit{pullpush.io}---a freely accessible clone of the PushShift API. 
This API facilitated the retrieval of Reddit posts, comments, and associated upvote and downvote counts, along with metadata for each post/comment, including timestamps, post/ comment/user identifiers, etx. The dataset comprises discussion threads from \swatch{} spanning the period from May 2023 to February 2024.
To ensure data integrity and respect user privacy, duplicate entries were removed, and any posts deleted by the user or removed by moderators were excluded.
In conclusion, our dataset includes 59,607 posts and 149,144 comments, averaging 2.64 comments per post. It contains 59,408 unique Reddit users who participated in \swatch{}.
\autoref{table:swatch} shows the descriptive statistics of our dataset.


\begin{table}[t!]
\centering
\sffamily
\caption{Descriptive statistics of our dataset from \swatch}
\footnotesize
\setlength{\tabcolsep}{2pt} 
\begin{tabular}{lr}
\toprule
Number of posts & 59,607\\
Number of users posting & 36,879\\
Avg. post length (words) & 150.71\\
Stdev. post length & 171.69\\
\hdashline
Number of comments & 149,144\\
Number of users commenting & 37,751\\
Avg. no. of comments per post & 2.64\\
Stdev. comments & 8.69 \\
Avg. comment length (words) & 22.43\\
Stdev. comment length & 38.18\\

\bottomrule
\end{tabular}
\label{table:swatch}
\end{table}

%% file: 4Aim1.tex
\section{Aim 1: Theory-driven Characterization of Suicidal Ideation in Online Spaces}

To address Aim 1, our study investigated how \sci{} manifests in online communities through the lens of the Interpersonal Theory of Suicide (\its{})~\cite{chu2017interpersonal,van2010interpersonal}. 
\its{} states that lethal suicidal behavior arises from the intersection of three \srsk{}s ---1) Thwarted Belongingness, 2) Perceived Burdensomeness, and 3) Acquired Capability for Suicide. Given its established relevance in understanding suicide risk, \its{} served as a theoretical framework in our work for categorizing expressions of \sci{} in the online community of \swatch{} on Reddit.

To operationalize \its{}, we employed a two-step approach where first we identified presence of \sdim{}s (Loneliness, Lack of reciprocal love, Self-hate, Liability) in the posts, and then based on the intersection of \sdim{}s, we labeled posts into the three \srsk{}s (Thwarted belonging, Perceived burdensomeness, and Acquired capability of suicide). Finally, from the intersection of \srsk{}s, we identified lethally suicidal posts---posts with the combination of all \its{} factors theorized to confer high risk of suicide attempts~\cite{van2010interpersonal}.

Our classification approach consisted of three major steps, 1) First, we classified posts in \sdim{}s by employing supervised learning trained on datasets resembling each \sdim{}s, 2) Second, we developed an iterative codebook-based similarities to refine the classification of posts into \sdim{}s, and 3) Third, we adopted a threshold-based intersection approach to label posts into \srsk{}s. This approach also enabled us to label posts exhibiting lethally suicidal ideation. We elaborate on our approach and validation in this section.

\subsection{Dimensions and Risk Factors of \sci{} as per \its{}}
\begin{figure}[t]
\centering
    \includegraphics[width=0.5\columnwidth]{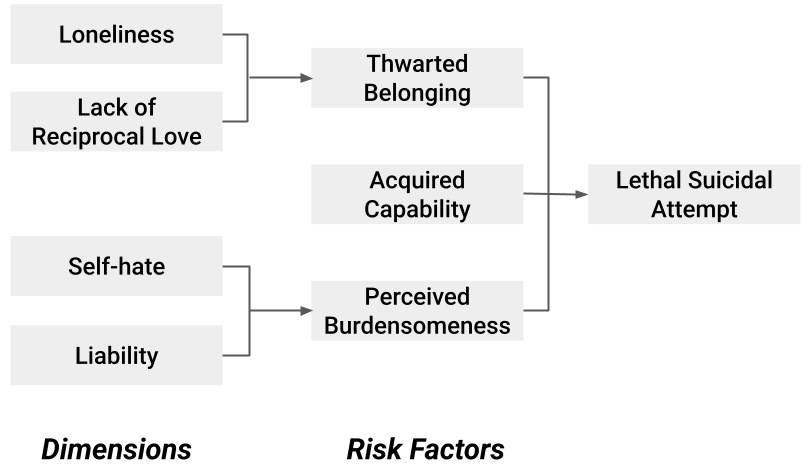}
\caption{A schematic representation on the causal pathway to lethal suicidal attempts based on the \its{}~\cite{van2010interpersonal}}
\label{fig:causal}
\end{figure}

To systematically classify posts based on the \its{}, we followed the causal pathway of \sci{} as outlined by ~\citeauthor{van2010interpersonal}, depicted in ~\autoref{fig:causal}. 
This framework enabled us to identify and label the \sdim{}s and \srsk{}s of \sci{} as per \its{}. We describe these below:

\para{Risk Factor: Thwarted belongingness}  
consists of the dimensions of 1) loneliness and 2) lack of reciprocal love---both of which are central to feelings of social alienation. 
Individuals experiencing thwarted belongingness often express distress related to social rejection, the absence of supportive relationships, or an enduring sense of being ignored or misunderstood.

\para{Risk Factor: Perceived burdensomeness}
consists of the dimensions of 1) self-hate and 2) liability (and self-guilt). 
Individuals experiencing this risk factor often express feelings of worthlessness, excessive guilt, or the perception that their presence negatively impacts family, friends, or society.

\para{Risk Factor: Acquired capability for suicide} is associated with a low fear of death and high pain tolerance, commonly associated with past experiences of self-harm or exposure to painful and traumatic events. Repeated exposure to painful experiences may desensitize individuals to the fear of death, thereby lowering the psychological barriers that typically deter suicidal actions.

\subsection{Classifying Posts into \its{} Types}\label{sec:its_classification}
To classify posts into the \srsk{}s and \sdim{}s as per \its{}, we adopted an iterative codebook-based similarity approach. 
This process was designed to systematically identify and categorize \sci{} expressions using a combination of distant supervision, semantic similarity matching, and iterative codebook refinement.
We began by classifying the \sdim{}s and Acquired Capability in \its{}, following the causal pathway in~\autoref{fig:causal}. 

\subsubsection{Distant supervision-based \sdim{} classification and identifying seed keywords}
First, we developed a distantly supervised binary classifier for each \its{} \sdim{}. 
These classifiers were trained using relevant distant datasets, as outlined below. 
The training process involved a 55:45 train-test split ratio. 
We trained each classifier model as a sequential neural network for binary text classification. 
This includes an embedding layer that converts words into 64-dimensional vectors. The model then includes an LSTM layer with 128 units, using dropout and recurrent dropout for regularization. A Global Max Pooling layer follows to capture the most significant features of the sequence. The model also has a dense layer with 64 units and rectified linear unit (ReLU) activation, followed by a sigmoid output layer for binary classification. It is compiled with binary cross-entropy loss and the Adam optimizer, with accuracy as the evaluation metric.

For \textbf{loneliness}, we used the expert-annotated \textit{LonelyDataset}~\cite{lonelydataset}, which comprises 5,633 text entries labeled as either ``lonely'' or ``not lonely.''
This dataset was selected due to its semantic alignment with Loneliness. For instance, the sentence ``I wish I could talk with someone'' is categorized as ``lonely,'' whereas ``I had a good conversation with dad'' is classified as ``not lonely''. 
On the test dataset, this classifier showed an accuracy of 93\%.

For \textbf{lack of reciprocal love}, we used the \textit{psychosocial-mental-health-analysis} dataset~\cite{love}. 
This dataset consists of posts classified into 30 categories, of which we focus on Relationships (37\%), Family (8\%), Interpersonal Conflicts (2\%), Divorce (1\%), and Marital Issues (1\%). 
An example entry from the ``Relationship'' category is, ``I experience feelings of loneliness, confusion, and mood swings, struggling to find fulfillment even when spending time alone,'' aligning directly with alack of reciprocal love. 
This classifier showed an accuracy of 95\%.

For \textbf{self-hate}, we used a validated hate speech dataset,\textit{measuring-hate-speech}~\cite{sachdeva2022measuring}. This dataset comprises 39,565 comments annotated by 7,912 annotators, resulting in 135,556 annotated instances. 
This dataset tags comments with a hate speech score, which we used for our study.
Although hate speech is not directly linked to self-hate, it contains keywords and key phrases associated with the broader theme of hate. 
The classification captures these patterns, providing a foundation for initiating the iterative codebook algorithm on our dataset. 

For \textbf{liability}, we used the \textit{LoST} dataset~\cite{garg2023lost}, which contains 3,252 text entries labeled as 0 (no low self-esteem or self-liability issues) or 1 (presence of such issues).. 
This dataset is especially relevant to identify liability, based on prior work that texts reflecting low self-esteem often convey perceptions of being a liability~\cite{teismann2024self}. 
This classfier showed a test accuracy of 87\%.

Accordingly, we constructed a comprehensive codebook of seed keywords and phrases (\autoref{table:seed_codebook}) that capture the language associated with each \sdim{} and acquired the capability for suicide. 
These key phrases are directly referred from the prior work~\cite{chu2017interpersonal,van2010interpersonal}. 

\begin{table}[t]
\centering
\sffamily
\footnotesize
\caption{A codebook of seed keywords for each secondary dimension and risk factors of of \sci{} as per \its{}.}
\begin{tabular}{p{0.12\columnwidth}p{0.85\columnwidth}}
\textbf{\its{}} & \textbf{Seed Keywords }\\
\toprule
\rowcollight \multicolumn{2}{l}{Risk Factor: Thwarted Belonging}\\
Loneliness & disconnected, loneliness, pulling together, no care, seasonal variation, reductions in social interactions, marriage, no children and friend, living alone, no social supports  \\
\hdashline
Lack of Reciprocal Love & lack love, no love, social withdrawal, low openness, single jail cell, domestic violence, childhood abuse, familial discord  \\
\hdashline
\rowcollight \multicolumn{2}{l}{Risk Factor: Perceived Burdensomeness}\\
Self-Hate &  I hate myself, I am useless, low self-esteem, self-blame, shame, mental state of agitation\\
\hdashline
Liability &  my death is worth more than my life, distress from homelessness, distress from incarceration, distress from unemployment, distress from physical illness, expendability unwanted, belief of burden on family\\
\hdashline
\rowcollight \multicolumn{2}{l}{Risk Factor: Acquired Capability}\\
Acquired Capability  &  increased physical pain tolerance, reduced fear of death, habituation, physical pain, acquired capability, lowered fear of death, past serious ideation, non-zero degree of fearlessness, courage and the ability to commit suicide, elevated physical pain tolerance, recent suicidal behavior, serious levels of \sci{}, cutting one's wrists, pulling the trigger on a gun, jumping off a building, overdose\\

\bottomrule
\end{tabular}
\label{table:seed_codebook}
\end{table}

\subsubsection{Labeling Posts with \its{} \sdim{}s}
We employed a semantic similarity approach, wherein each classified post from the earlier distantly supervised model was compared against the seed key phrases in the codebook (\autoref{table:seed_codebook}). Both posts and key phrases were embedded into a 384-dimensional vector space using 
a transformer-based language model of MiLM~\cite{wang2020minilm}.
We computed cosine similarity scores between the embedded representations of posts and \sdim{}'s codebook phrases, labeling a post under a specific \sdim{} if the similarity exceeds a threshold of 0.6. 
This threshold was determined through iterative experimentation of trial and error to optimize label quality for each post, and also aligns with prior research adopting similar methodologies of language embedding-based similarities~\cite{saha2018social,rekabsaz2017exploration}.

The labeling process was refined by iteratively updating the codebook. 
We used a Rapid Automatic Keyword Extraction (RAKE)~\cite{rose2010automatic} method to extract 
key phrases from the previously labeled posts and incorporated them into the codebook. 
We then re-applied the semantic similarity approach, repeating this process of extracting keywords and updating the codebook until the codebook remained constant. This iterative refinement ensured broader coverage of language patterns commonly associated with \sci{} within \its{} framework.

Upon labeling all the posts into \sdim{}s, we identified \srsk{}s by detecting posts containing the two associated \sdim{}s with a given \srsk{} (depicted in \autoref{fig:causal}). 
The cosine similarity scores of these \sdim{}s were averaged, and posts with scores above the threshold (0.6) were assigned to the corresponding \srsk{}. 
This multi-dimensional labeling approach allowed for a nuanced understanding of \sci{}, highlighting the interrelated nature of different \srsk{}s.


\subsubsection{Validating the \its{} Classification} We evaluated the dataset for the initial distantly supervised binary classifier model using a linguistic equivalence test inspired by prior work~\cite{saha2019social}. We measured word embedding-based similarities between the posts classified under each \sdim{} and corresponding semantically relevant datasets to understand if the dataset used was the correct dataset to train such a classifier model or not.

The results demonstrated high similarity scores, indicating strong alignment between distant data and classified \sdim{} posts of \sci{}---80.4\% for loneliness, 83.3\% for lack of reciprocal love, 94.3\% for self-hate, and 76.2\% for liability. 
These results reinforced the effectiveness of the semantically relevant datasets used to train the distantly supervised binary classifier, capturing the linguistic characteristics of each \sdim{}.

Furthermore, to understand the final classification of posts in the four \srsk{}, we conducted an expert validation of 450 classified posts.
For \sdim{}s, we selected 50 posts each for loneliness, lack of reciprocal love, self-hate, and liability. 
For \srsk{}s, we examined 50 posts for thwarted belongingness, perceived burdensomeness, acquired capability for suicide, and finally, 50 posts for lethally suicidal tendencies.
This dataset sample was manually verified by one coauthor with long-term experience in social media analyses and mental health disclosures.
Specifically, we adopted the underlying theoretical framework of \its{} to manually label \its{} \sdim{}s and \srsk{}s in the small sample of posts.
We found a high level of agreement between manual assessments and our \its{} labeling. Among \sdim{} labels, the automated and human ratings showed agreement rates of 93\% for loneliness, 94\% for lack of reciprocal love, 94\% for self-hate, and 88\% for liability. 
Similarly, for \srsk{} labels, match rates were 88\% for thwarted belongingness, 88\% for perceived burdensomeness, 80\% for acquired capability for suicide, and 74\% for lethally suicidal posts. 
The manual evaluation provides stronger reliability of our classification approach and highlights its potential for identifying \sci{} in online spaces with a high degree of accuracy. 
These findings further support the applicability of \its{} in online spaces of \sci{} disclosures.

\subsubsection{Distribution of \sci{} Disclosures}
Using our computational approach discussed above, we labeled our entire dataset of $\sim$59K posts with \sci{} \sdim{} and \srsk{}s. 
\autoref{table:dimension_distr} presents the distribution of these labels within our dataset.
The most frequently expressed \sdim{} was loneliness, accounting for 24\% of posts (12,091 instances), indicating its predominant role in online \sci{} discourse. Among the three \srsk{}, thwarted belongingness appeared in 16\% of posts (8,171 instances), making it the most prevalent \srsk{}. This was followed by perceived burdensomeness, which was present in 5.8\% posts (3,441 instances) and acquired capability for suicide, which was identified in 3.3\% of posts (1,980 instances).

Notably, 2.5\% of posts (1,508 instances) exhibited all three \srsk{}s, classifying them as lethally suicidal. 
These posts exhibited the highest risk level, as individuals expressing all three \srsk{}s
are considered to be at an elevated risk for suicidal behaviors, according to the \its{}. 
To provide further context,~\autoref{post_example} presents example paraphrased posts corresponding to each \its{} \sdim{} and \srsk{}. These examples illustrate how \sci{} manifests in online discussions. Additionally, post-hoc qualitative explanations are provided for each classification, ensuring transparency in the interpretability of the model’s decision-making process.

\begin{table}[t!]
\centering
\sffamily
\caption{Distribution of the posts by \its{} dimensions. A post is classified as a \srsk{} if it shows a similarity to both its corresponding \sdim{}s is greater than 0.60 (Acquired capability is classified directly).}
\footnotesize
\setlength{\tabcolsep}{2pt} 
\begin{tabular}{lrl}
\textbf{\its{} Type} & \textbf{Number of posts}\\
\toprule
Risk Factor: Thwarted Belonging & 8,171 & \dgreybar{8.171}\\
~~Dimension: Loneliness & 12,091 & \dgreybar{12.091} \\
~~Dimension: Lack of Reciprocal Love & 11,422  & \dgreybar{11.422}\\
\hdashline
Risk Factor: Perceived Burdensomeness & 3,441 & \dgreybar{3.441}\\
~~Dimension: Self-Hate &  9,760 & \dgreybar{9.760}\\
~~Dimension: Liability &  10,141 & \dgreybar{10.141}\\
\hdashline
Risk Factor: Acquired Capability & 1,980 & \dgreybar{1.980}\\
\rowcollight Lethally Suicidal & 1,508 & \dgreybar{1.508}\\
\bottomrule
\end{tabular}
\label{table:dimension_distr}
\end{table}


\begin{table}[t]
\sffamily
\footnotesize
\centering
\setlength{\tabcolsep}{3pt} 
\caption{Examples of (paraphrased) posts for each \sdim and \srsk as per \its, along with explanations.}
\begin{tabular}{p{0.11\textwidth}
                p{0.55\textwidth}
                p{0.3\textwidth}}

\textbf{Dimension/ Risk Factor} & \textbf{Example Post} & \textbf{Explanation} \\
\toprule
Dimension: Loneliness & I'm lost and feel empty without purpose or joy. I used to have a partner who made life meaningful, but now everything feels pointless. [..] Maybe she left because I wasn't going anywhere. Even though no one reads this, I still write my thoughts because I'm too afraid to end it all. I don't think I have it in me to change. & This post reveals deep feelings of loneliness, where the author is isolated and overwhelmed by emotional pain following the loss of a partner. \\

\rowcollight Dimension:\newline Lack of Reciprocal Love & I have a good life but struggle with low self-esteem and can't understand how anyone, including my brothers, could love me. [..] Despite putting my pain into art, it’s not enough, and I often visualize harming myself. [..] I want to live, but not like this, and I can't afford therapy or ask my mother for help again. & Here, the individual craves affection and validation but feels unworthy of love. This emotional disconnect deepens feelings of isolation, reinforcing a belief that love is conditional or unattainable. \\

Dimension:\newline Self-Hate & I'm giving it until February to decide if I should stay. After years of struggling with mental health, [..] I despise myself, my body, and my personality. [..] After a sexual assault in March, I feel like everything is my fault. Despite seeking help, nothing worked, and I no longer want to improve. & Reflects deep self-hatred stemming from mental health struggles and trauma. Feelings of worthlessness, guilt, and failed attempts to improve lead to a belief that they are a burden hold no value.\\

\rowcollight Dimension:\newline Liability & I’m 17 and all I want is to die. [...] My health is a mess—I don’t exercise, my sleep is terrible, and I’m always tired. I have to rely on my parents for money, which makes me feel guilty. I know something’s wrong with me, but I have no idea how to fix it. & Conveys a belief that achievements are worthless due to financial struggles, health issues, or personal dissatisfaction. Feels unable to meet societal expectations, leading to guilt and a sense of letting others down. \\

Risk Factor: Acquiring Capability for Suicide & I’ve made up my mind about killing myself. [...] I’ve been cutting myself more and trying riskier things, getting comfortable with the idea of slitting my wrists or overdosing. I’m going to die. & Reveal a deep focus on suicide, with many expressing intent to end their lives and referencing past struggles with suicidal thoughts and self-harm. \\

\rowcollight Risk Factor:\newline Thwarted Belonging & I feel like a real-life NPC, unable to connect with my life despite having a stable living situation. Childhood trauma, particularly related to homophobia, may contribute to my emotional numbness. [..] Despite living in a supportive environment my life feels pointless and detached. & Convey a sense of deep emotional struggles, isolation, and detachment from life. \\

Risk Factor:\newline Feeling of Burdensomeness & I feel useless and hypocritical, struggling with anxiety and depression. I have breakdowns and suicidal episodes but feel guilty for asking for help. I resort to cutting my forearms to cope, torn between wanting to end my life and longing for happiness. & Conveys deep feelings of worthlessness, guilt, and isolation. Individuals feel like a burden to their loved ones, intensifying their emotional pain and hopelessness. \\

\rowcollight Risk Factor:\newline Lethal Suicidal Attempts & I need help overcoming my fear of suicide and death because I've attempted suicide twice but hesitated both times. I feel like a fraud compared to those who go through with it, and when I tried not to eat as a suicide attempt, I failed. & Reveal deep struggle with hopelessness, worthlessness, and isolation, often with a history of suicide attempts. They experience chronic loneliness and emotional disconnection from others.\\
\bottomrule

\end{tabular}
\label{post_example}
\end{table}

\subsection{Characterizing Posts by \its{} Types}
\begin{figure}[t]
\centering
    \includegraphics[width=0.5\columnwidth]{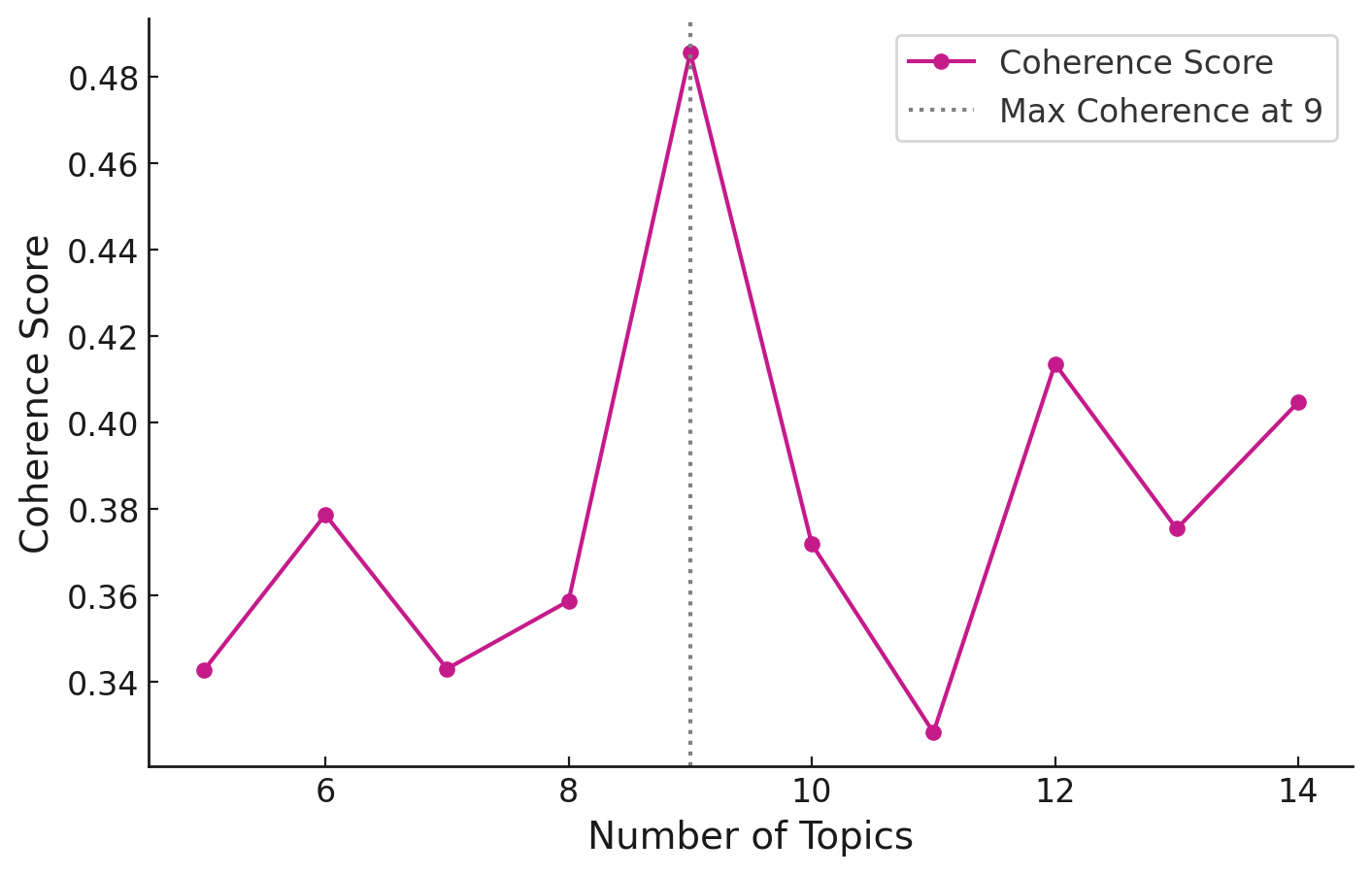}
    \captionof{figure}{Coherence scores by varying the number of topics ($k$) in BERTopics topic model.}
    \label{fig:coherence_scores}    
\end{figure}

We analyzed the language of \srsk{}s and \sdim{}s using topic modeling.
For topic modeling, we employed BERTopic models~\cite{grootendorst2022bertopic,saha2025mental}, followed by manual inspection and labeling of topics.
In BERTopic, we varied the number of topics ($k$) ranging between 5 and 14 (ref:~\autoref{fig:coherence_scores})---we found a highest coherence score at $k$=9. 
We manually inspected the topical clusters, and labeled these topics. 
Out of these 9 topics, we dropped Topic -1, which consisted of outlier words, and could not be assigned a meaningful label. 
This led to our eight final topics, as summarized in~\autoref{table:topics}, along with explanation and top keywords. 

Then, we obtained the distribution of these topics within our dataset, and mapped the topical occurrences with \srsk{}s and \sdim{}s, as presented in~\autoref{theme_distribution}.
The theme ``General Despair and Emotional Struggle'' is prominent across all \sdim{}s, \srsk{}s, and lethally suicidal posts. This suggests that individuals in emotional distress face multiple suicidal factors. It is particularly strong in the Lack of Love and Loneliness dimensions, indicating higher emotional pain in those feeling disconnected.


\begin{table}[t]
\centering
\footnotesize
\sffamily
\setlength{\tabcolsep}{2pt} 
\caption{Topics identified as per BERTopics with explanations and top keywords.}
\begin{tabular}{p{0.2\textwidth}p{0.38\textwidth}p{0.38\textwidth}}

 \textbf{Topic Theme} & \textbf{Explanation} & \textbf{Keywords} \\
\toprule
Despair and Emotional Struggles & Expressions of hopelessness, dissatisfaction, and mental turmoil & don't want life, feel like, think, anymore, lost and broken, no hope left \\

\rowcollight Substance Use & Substances like ibuprofen or household items, potentially linked to self-harm or risky behaviors & hand\_soap, drank, ibuprofen advil, advil kinda funny, mixing alcohol, took pills again \\

Seeking Support or Validation & Highlights the need for emotional support, or recognition, suggesting a plea for interaction & need someone, need, talk, guess who back, good person, I need help, someone please listen\\

\rowcollight Weakness and Pain & Feelings of physical and emotional exhaustion, feeling broken, and recalling painful experiences & damaged weak, damaged, weak, tired, hurt, fair, worst days, always broken\\

Planning and Attempts & Thoughts or actions related to planning self-harm, including specific methods or failed attempts & bag tied around, tied around, xannns, failed, head, took, writing a note, ready to jump\\

\rowcollight Non-suicidal self-injury (NSSI) & Focuses on self-inflicted harm and the immediate consequences & currently bleeding everywhere, bleeding, finally,can’t stop cutting, razor in hand\\

Methods and Tools & Discusses methods involving specific tools or materials, possibly related to planned attempts & helium tanks, air would, oxygen mask, asphyxiation, found the rope, sharp enough blade \\

\rowcollight Cynicism and Bitterness & Expressions of cynicism and dark humor as a coping mechanism, using humor to mask pain with life & haha stay low, assholes haha, nothing ever changes, it’s all a joke, world’s full of idiots \\

\bottomrule
\end{tabular}
\label{table:topics}
\end{table}



Another significant theme, ``Seeking Support or Validation'', appeared across all \srsk{} and \sdim{}. This suggests that individuals experiencing distress often reach out to online communities, possibly searching for connection, reassurance, or understanding. Its strong presence in the Lack of Love, Loneliness, and Self-Hate dimensions highlights that such individuals feel deeply alienated craving for any sort of validation to reduce their \sci{}. 
This aligns with prior work that found online help-seekers often experience higher levels of psychological distress and are more likely to seek anonymity due to stigma~\cite{wong2021seeks}.

The ``Feelings of Weakness and Pain'' theme showed a strong association with Lethally Suicidal posts and Acquired Capability of Suicide. This correlation highlights the link between repeated suicidal attempts and increased pain tolerance, suggesting that prolonged distress and self-harm contribute to a heightened risk of suicidal behavior. Similarly, the ``Non-suicidal self-injury (NSSI)'' theme was strongly associated with Lethally Suicidal and Acquired Capability, reinforcing the notion that individuals with a history of self-harm tend to have an increased tolerance for pain, making them more susceptible to considering or attempting suicide. This theme also appeared in the Self-Hate and Lack of Love dimensions, suggesting that negative self-perceptions and emotional deprivation contribute to self-destructive tendencies. This finding is also stated in \its{}~\cite{joiner2010interpersonal}.

Additionally, ``Cynicism and Bitterness'' was highly prevalent in the Loneliness dimension. This indicates that individuals who feel socially disconnected and emotionally isolated may develop a cynical worldview, potentially worsening their emotional state and reducing their willingness to seek help. 
This aligns with prior work that link social isolation to increased negativity and reduced engagement in support-seeking behaviors~\cite{brandt2022effects}.

A particularly concerning theme, ``Planning and Attempts'', was strongly associated with the \srsk{} of Acquired Capability. This suggests that individuals who actively plan suicidal attempts often reach a critical threshold of pain tolerance. This theme indicates a transition from ideation to intent, marking a dangerous phase where individuals not only contemplate death but also take steps toward enacting it~\cite{joiner2010interpersonal, chu2017interpersonal, van2010interpersonal}. 

Overall, the presence of themes like General Despair, Seeking Support, and Planning and Attempts highlights both the urgency of intervention and the potential role of online support in shaping suicidal trajectories.

\begin{table}[t]
\centering
\sffamily
\footnotesize
\setlength{\tabcolsep}{2pt} 
\caption{Normalized distribution of BERT topics themes across \sdim{} and \srsk{} on posts}
\resizebox{\columnwidth}{!}{\begin{tabular}
{lp{0.09\textwidth}p{0.09\textwidth}p{0.09\textwidth}p{0.09\textwidth}p{0.09\textwidth}p{0.09\textwidth}p{0.09\textwidth}p{0.09\textwidth}}
\multirow{2}{*}{\textbf{Topic Theme}} & \multicolumn{4}{c}{\textbf{\srsk{}}} & \multicolumn{4}{c}{\textbf{\sdim{}}} \\
\cmidrule{2-9}
& \textbf{Lethally\newline Suicidal} & \textbf{Thwarted Belonging} & \textbf{Perceived Burden} & \textbf{Acquired Capability} & \textbf{Lack of\newline Love} & \textbf{Loneliness} & \textbf{Self Hate} & \textbf{Liability} \\
\toprule
Despair and Emotional Struggle & 0.0119 & 0.0059 & 0.0145 & 0.0106 & 0.0061 & 0.0017 & 0.0097 & 0.0034\\
\rowcollight Substance Use & 0.7016 & 0.6816 & 0.7167 & 0.6513 & 0.7147 & 0.1120 & 0.0774 & 0.0815\\
Seeking Support or Validation & 0.1379 & 0.1576 & 0.1363 & 0.1524 & 0.1299 & 0.0480 & 0.0102 & 0.0226\\
\rowcollight Weakness and Pain & 0.0007 & 0.0001 & 0.0003 & 0.0005 & 0.0002 & 0.0002 & 0.0003 & 0.0001\\
Planning and Attempts & 0.0711 & 0.0570 & 0.0552 & 0.0815 & 0.0697 & 0.0186 & 0.0161 & 0.0286\\
\rowcollight Non-suicidal self-injury (NSSI) & 0.0029 & 0.0027 & 0.0021 & 0.0027 & 0.0023 & 0.0014 & 0.0020 & 0.0011\\
Methods and Tools & 0.0895 & 0.0793 & 0.0706 & 0.0989 & 0.0718 & 0.0157 & 0.0056 & 0.0089\\
\rowcollight Cynicism and Bitterness & 0.0040 & 0.0026 & 0.0058 & 0.0069 & 0.0060 & 0.0072 & 0.0035 & 0.0006\\
\bottomrule
\end{tabular}}
\label{theme_distribution}
\end{table}

%% file: 5Aim2.tex
\section{Aim 2: Analyzing Language of Responses to \sci{}  in Online Spaces}
From Aim 1, we gained a deeper understanding of the nuanced characteristics that define \sci{} posts. Now, in this section, we aim to explore how community members respond to such SI posts.
For this purpose, we examined the language of responses (comments) to \sci{} posts using, 1) Psycholinguistics lexicon, Linguistic Inquiry and Word Count (LIWC)~\cite{tausczik2010psychological}, and 2) Content analysis, Sparse Additive Generative Model (SAGE)~\cite{eisenstein2011sparse}. 
This dual approach of linguistic analysis and keyword differentiation provided a robust framework for understanding the nuanced language and psychological profiles associated with the responses to each risk factor.

\subsection{Psycholinguistic Analysis of Comments - LIWC}
We analyzed the comments' emotional tone, cognitive processes, social concerns, and other psychological factors using the LIWC tool~\cite{francis1993linguistic}. Linguistic and emotional markers were identified by comparing normalized LIWC category occurrences across the three \its{} \srsk{}s. \autoref{table:liwc} shows the distribution of these categories and the Kruskal-Wallis H-test for statistical significance. Our findings are detailed below:

\begin{table*}[t]
\centering
\sffamily
\footnotesize
\caption{Normalized occurrences of psycholinguistic markers in the comments to \sci{} posts varying in risk factors across thwarted belongingness, perceived burdensomeness, and acquired capability, along with Kruskal-Wallis $H$-stat. Statistical significance reported after Bonferroni correction (* p\textless{}0.05, ** p\textless{}0.01, *** p\textless{}0.001).} 

\begin{minipage}[t!]{0.5\columnwidth}
\centering
\resizebox{\columnwidth}{!}{
\begin{tabular}{lrrrrr@{}lr@{}l}
\textbf{LIWC} & \textbf{Thwarted B.} & \textbf{P. Burden.} & \textbf{Acquired C.} & \textbf{H-stat.} & \\ 
\toprule
\rowcollight \multicolumn{9}{l}{\textbf{Affect}}\\
Pos. Affect & 0.133 & 0.136 & 0.133 & 66.068***\\
Neg. Affect & 0.037 & 0.033 & 0.038 & 319.863***\\
Anxiety & 0.003 & 0.004 & 0.003 & 346.375***\\
\hdashline
Anger & 0.009 & 0.008 & 0.008 & 244.097***\\
Sad & 0.011 & 0.008 & 0.012 & 681.115***\\

\rowcollight \multicolumn{9}{l}{\textbf{Cognition and Perception}}\\
Insight & 0.034 & 0.029 & 0.031 & 443.637***\\
Causation & 0.017 & 0.017 & 0.019 & 195.374***\\
\hdashline
Tentat. & 0.097 & 0.104 & 0.085 & 799.830***\\
Certainty & 0.025 & 0.023 & 0.025 & 338.564***\\
Differ. & 0.099 & 0.097 & 0.095 & 84.117***\\
\hdashline
See & 0.013 & 0.012 & 0.012 & 661.096***\\
Hear & 0.016 & 0.022 & 0.018 & 165.520***\\
Feel & 0.017 & 0.014 & 0.017 & 281.563***\\

\rowcollight \multicolumn{9}{l}{\textbf{Biological Processes}}\\
Body & 0.012 & 0.011 & 0.014 & 617.260***\\
Health & 0.028 & 0.025 & 0.032 & 49.545***\\
Sexual & 0.008 & 0.008 & 0.006 & 408.176***\\

\rowcollight \multicolumn{9}{l}{\textbf{Informal}}\\
Informal & 1.020 & 1.025 & 0.977 & 735.52***\\
Swear & 0.016 & 0.013 & 0.014 & 654.62***\\
Assent & 0.057 & 0.066 & 0.063 & 44.59***\\
\hdashline
Non-flu. & 0.073 & 0.083 & 0.068 & 653.31\\

\bottomrule
\end{tabular}}
\end{minipage}\hfill
\begin{minipage}[t!]{0.5\columnwidth}
\centering
\resizebox{\columnwidth}{!}{
\begin{tabular}{lrrrrr@{}lr@{}l}
\textbf{LIWC} & \textbf{Thwarted B.} & \textbf{P. Burden.} & \textbf{Acquired C.} & \textbf{H-stat.} & \\ 
\toprule
\rowcollight \multicolumn{9}{l}{\textbf{Social \& Personal Concerns}}\\
Family & 0.029 & 0.030 & 0.033 & 10.95***\\
Friends & 0.011 & 0.014 & 0.011 & 630.29***\\
Leisure & 0.013 & 0.016 & 0.015 & 51.66***\\
\hdashline
Home & 0.004 & 0.003 & 0.004 & 151.97***\\
Religion & 0.004 & 0.004 & 0.005 & 236.08***\\
\hdashline
Space & 0.198 & 0.195 & 0.198 & 5.08\\
Time & 0.073 & 0.069 & 0.066 & 506.88***\\
Achievement & 0.023 & 0.023 & 0.021 & 567.63***\\
\hdashline
Power & 0.039 & 0.029 & 0.038 & 318.61***\\

\rowcollight \multicolumn{9}{l}{\textbf{Function Words}}\\
Preposition & 0.283 & 0.285 & 0.282 & 10.97**\\
Conjunction & 0.147 & 0.142 & 0.141 & 63.01***\\
Adverb & 0.099 & 0.102 & 0.105 & 9.37**\\
\hdashline
Negation & 0.037 & 0.033 & 0.044 & 78.90***\\
Aux. Verb & 0.195 & 0.197 & 0.199 & 5.39\\
Verb & 0.344 & 0.345 & 0.358 & 308.60***\\
\hdashline
Adjective & 0.105 & 0.099 & 0.109 & 43.15***\\
Compare & 0.048 & 0.050 & 0.054 & 2.83\\
Number & 0.016 & 0.011 & 0.015 & 477.93***\\
\hdashline
Quantifier & 0.044 & 0.048 & 0.049 & 14.19**\\

\rowcollight \multicolumn{9}{l}{\textbf{Temporal References}}\\
Past & 0.042 & 0.035 & 0.044 & 145.914***\\
Present & 0.302 & 0.311 & 0.316 & 204.571***\\
Future & 0.022 & 0.021 & 0.023 & 103.121***\\

\bottomrule
\end{tabular}}
\end{minipage}

\label{table:liwc}
\end{table*}

\para{Affect} category reflects the emotional tone, capturing positive and negative emotions. 
Positive affect was highest for the responses to perceived burdensomeness risk factor (0.136), showing that respondents used more positive language to such individuals. 
Negative affect was highest in responses to thwarted belonging (0.37), indicating more negative emotions while responding to those feeling socially isolated, such as in, ``I'm feeling in a suicidal state [..] In constant pain, and tired with zero libido, so my sex life has been wrecked too [..] Suicidal thoughts are constant here.'' 
Responses to acquired capability exhibited a higher likelihood of sadness, as seen in a comment, ``I'm very sorry all of that happened. Hope you can find hope and light in the darkness.'' In this case, the responder expressed sorrow and offered sympathy to an individual who had previously attempted suicide.

\para{Cognition and Perception} attribute include thought processes and perception-related words.
Responses to perceived burdensomeness showed high usage of tentative language (0.104), indicating uncertainty, as in,
``I get where you’re coming from [..] Yet because I understand doesn't make it any less hurtful''.
Responses to acquired capability also showed higher expressions of feeling (0.009), focusing on pain and suffering-related words.

\para{Biological Processes} cover bodily states, health, and physical sensations. 
Responses to the acquired capability of suicide mentioned the most about health (0.032), indicating physical concerns.
For example, a commenter responded about their personal experience of coping with extreme sexual desires, how harming oneself is not the answer, and urged to search for healthier alternatives.
Responses to thwarted belonging included words related to sexuality (0.008), suggesting romantic disconnection, and responses to Acquired capability showed high occurrences of body-related terms (0.014), focusing on preventing self-inflicted body harm, as in, ``You need to fight. Yes, I'm talking to you[..] I will never do anything to cut myself [..]''

\para{Social and Personal} attributes consists words on social connections, family, and personal interests.
Responses to perceived burdensomeness focused more on friends (0.014), indicating concerns about how it affected relationships and social life, such as in, ``I’ve opened up before, but it was used against me [..] Part of me misses the past relationships, but another part just wants a friend without feeling like a burden.'' 
Acquired capability responses mentioned religion (0.005), and power (0.038) as seen in, ``During my time in the psych ward, a therapist suggested I find a religion [..] 
After learning about the tenets and beliefs,, I decided to become a Satanist''.
Also, responses to thwarted belonging showed references to achievements (0.023) and time (0.073), reminding the individual of their accomplishments, helping them feel valued, and giving them a sense of belonging. This is seen in the example, ``You're doing your best, I can see that [..] It takes so much to get by every single day in your condition.''

\para{Function Words} focus on language structure, such as articles, conjunctions, and prepositions. 
Responses to acquired capability are likely to use more verbs (0.358), adjectives (0.109), and negation (0.044), reflecting emotionally intense efforts by the responders to stop the individual from further harming themselves as seen in, ``[..] You're so young! [..] The only thing that's stopping me from killing myself is the idea of another failure like, what if I failed to kill myself and end up waking in a hospital''.

\para{Temporal References} consist of time-related words, indicating thoughts' orientation in time. 
For Temporal References, acquired capability comments focused more on the past (0.044), present (0.316), and future (0.023), reflecting attention to their past, current, and future states trying to connect with the individual who posted originally. 
For example, one \oc{} member narrated their personal experience, ``I don’t tell people it will get better. The only thing I say is that things have gotten better for me since my last attempt [..]''.

\begin{table}[t]
\sffamily
\footnotesize
\centering
\caption{Top discriminating $n$-grams ($n$=1,2) in responses to lethal and non-lethal \sci{} posts with SAGE~\cite{eisenstein2011sparse}. Positive SAGE indicate saliency in responses to lethal \sci{} posts and negative SAGE indicate saliency in responses to non-lethal \sci{} posts.}

\begin{minipage}[t]{0.22\columnwidth}
\centering
\begin{tabular}{lr}
\textbf{$n$-gram} & \textbf{SAGE} \\
\toprule
life problems & 5.64\\ 
strange place & 5.64\\
change lot & 5.64 \\
probiotics & 5.41 \\
know ever & 5.41  \\
like partner & 5.41 \\
go pain & 5.41 \\
grow apart & 5.41 \\
\bottomrule
\end{tabular}
\end{minipage}\hfill
\begin{minipage}[t]{0.22\columnwidth}
\centering
\begin{tabular}{lr}
\textbf{$n$-gram} & \textbf{SAGE} \\
\toprule
many problems & 5.41 \\
nothing good & 5.41\\
often times & 5.41\\ 
learn write & 5.41\\
cents & 5.41 \\ 
mid-twenties & 5.41\\
crimes & 5.41 \\
matter small & 5.10 \\
\bottomrule
\end{tabular}
\end{minipage}\hfill
\begin{minipage}[t!]{0.22\columnwidth}
\centering
\begin{tabular}{lr}
\textbf{$n$-gram} & \textbf{SAGE} \\
\toprule
penny & -5.08 \\ 
 thats okay & -4.67 \\
intrusive & -4.45 \\
 intrusive thoughts & -4.44 \\
theyve & -4.39 \\
 bad thing & -4.22 \\
views & -4.20\\
 find penny & -4.18 \\
\bottomrule
\end{tabular}
\end{minipage}\hfill
\begin{minipage}[t!]{0.22\columnwidth}
\centering
\begin{tabular}{lr}
\textbf{$n$-gram} & \textbf{SAGE} \\
\toprule
friend like & -4.17 \\
 caffeine & -4.12 \\
coworker & -4.11 \\ 
 small moments & -4.11 \\
karma would & -4.10 \\
everywhere & -4.02 \\
go get & -4.02 \\
sexual life & -4.00 \\
\bottomrule
\end{tabular}
\end{minipage}
\label{table:sage_lethal}
\end{table}

\subsection{Content Analysis of Responses to Lethal vs. Non-Lethal \sci{} Posts}
We examined content differences in responses to lethal (intersection of all three risk factors) versus non-lethal \sci{} posts. 
To identify differences in responses, we employed the Sparse Additive Generative Model (SAGE)~\cite{eisenstein2011sparse}. SAGE compares parameters of two logistically parameterized multinomial models, with a self-tuned regularization parameter to balance frequent and rare terms. 
We applied SAGE to identify distinguishing $n$-grams ($n$=1,2) between responses to lethal and non-lethal \sci{} posts. The SAGE magnitude captures uniqueness—positive values indicate terms more likely in responses to lethal posts, negative values in non-lethal ones. 
\autoref{table:sage_lethal} summarizes distinguishing keywords in responses to lethal and non-lethal \sci{} posts. 

Responses to lethal \sci{} posts showed keywords on pain and life, suggesting painful issues in life. Phrases like \textit{life problems} and \textit{many problems} highlighted overwhelming hardships, while \textit{grow apart} suggested isolation. \textit{Nothing good} indicated a bleak outlook, and \textit{body real} and \textit{crimes} suggested victimization, likely by harming one's body. 
We found a distinct pattern where commenters expressed pessimism and used depressive language, suggesting a tendency to relate the original post to personal experiences. 
For example, words such as \textit{death} and \textit{strange}, pointed to exhaustion, with suicide seeming like the only escape, which could be seen in, ``[..] I’m under the firm belief that life is not worth it and that death is good [..]''

In responses to non-lethal \sci{} posts, keywords like \textit{intrusive thoughts} and \textit{bad thing} reflected negative self-perceptions and feelings of being a burden. \textit{Friend like} and \textit{coworker} suggested superficial connections and disconnection, as seen in, ``Realizing how far my avoidant tendencies go, pushing away from all my friend-like connections [..] Actively destroying my friendship over stupid stuff''. 
We also found occurrences of keywords such as \textit{small moments}, which indicated assistance in helping to find joy, whereas \textit{tell someone} indicated a desire to help individuals on the platform having \sci{}.

Together, this section analyzed the language used in responses to \sci{} posts, which revealed insights into the emotional and psychological states of individuals responding. 
We found, people respond differently to different \sci{} posts. 
When the individual felt like a burden, responses were more positive but contained uncertainty. When someone felt isolated, responses were the most negative and also uncertain, reflecting doubts about unhealthy social connections. Responses to those who had attempted or thought about suicide focused more on health and power, using higher verbs and strong emotions, denoting urgent language to discourage self-harm. These responses also referenced the past, present, and future, helping build connections through personal stories. This was also observed by looking at the SAGE-based content analysis between responses to lethal and non-lethal \sci{} posts, where responses to lethal \sci{} posts share their own suicide attempt story. 

%% file: 6Aim3.tex
\section{Aim 3: Evaluating AI's Responses to Suicidal Ideation}



The advancements in LLMs have enabled high-quality, natural language responses to user queries. 
AI chatbots present a potential approach for delivering timely and effective supportive responses to posts on \sci{}. 
Therefore, for Aim 3, we examined how an AI would respond to online mental health queries on \sci{}. 
We explored whether prompting with linguistic cues of supportiveness could enhance the quality of the AI chatbot's responses. 

\subsection{Generating AI Responses}
We prompted a state-of-the-art LLM as our AI chatbot, GPT-4o, with varying levels of context (including \its{} categories), and conducted a linguistic comparison on lexico-semantic attributes.
We conducted our analyses by prompting with three kinds of context settings:

\para{AI-1: Prompting only posts.} In this setting, we prompted the Reddit post. 
This is more of a baseline scenario of AI responses in terms of how the model interprets and responds to \sci{} posts. 

\para{AI-2: Prompting posts and \its{} category.}
In this setting, we prompted the post along with its \its{} category based on our classification of the \sci{} post (from Aim 1).


\para{AI-3: Prompting posts, \its{} category, and linguistic characteristics.} In this setting, we prompted the AI with the post, \its{} category, as well as key features of supportive responses as per prior literature~\cite{saha2020causal}. These characteristics include that the response should be 1) semantically similar and linguistically accommodating to the query, 2) diverse, 3) empathetic, and 4) promoting hopefulness. 

Our tiered approach of prompting was aimed to offer a systematic evaluation of how contextual enrichment influenced the response quality of the AI chatbot.
We obtained a random sample of 2,000 posts from our dataset---where 500 posts exhibited each of the three \srsk{}s and 500 exhibited lethally \sci{}, and then prompted these posts to GPT-4o using the above settings of prompts. 

\subsection{Comparing AI and Human-written Responses}

After generating AI responses across various contextual settings, we conducted a comparative analysis against human-written responses from online communities (\oc{}). 
For this purpose, we built upon previous research and performed a comprehensive suite of lexico-semantic analyses~\cite{saha2020causal,althoff2016large,saha2025ai,dasswain2025ai,saha2025linguistic}.
\autoref{ai_compare} presents an overview of these comparisons.
For each metric, we conducted $t$-test to compare each type of AI response with \oc{} responses and a Kruskal-Wallis $H$-test across all responses to assess statistical significance.
The following paragraphs elaborate on the operationalization of these lexico-semantic metrics and key observations derived from the analysis.

\begin{table}[t!]
\centering
\sffamily
\footnotesize
\caption{\textbf{All Risk Factors}: Summary of comparing the responses on online communities (OC) and by multiple AI---AI-1 (default GPT-4), AI-2 (GPT-4 with Themes), and AI-3 (GPT-4 with Themes and Characteristics of Supportive Responses), including paired $t$-tests in comparison with OC responses, and a Kruskal-Wallis $H$-test across all the four modalities ($ * p <0.05, ** p<0.01, *** p<0.001$).}

\begin{tabular}{lrrr@{}lrr@{}lrr@{}lr@{}l}
\setlength{\tabcolsep}{1pt}\\
& \textbf{OC} & \multicolumn{3}{c}{\textbf{AI-1}}& \multicolumn{3}{c}{\textbf{AI-2}}& \multicolumn{3}{c}{\textbf{AI-3}} & & \\ 
\cmidrule(lr){2-2}\cmidrule(lr){3-5}\cmidrule(lr){6-8}\cmidrule(lr){9-11}
\textbf{Categories}  & \textbf{Mean} & \textbf{Mean} &  \textbf{t-test} &  & \textbf{Mean} & \textbf{t-test} &  & \textbf{Mean} & \textbf{t-test} &  & \textbf{H-stat.}& \\ 
\toprule

\rowcollight \multicolumn{13}{l}{Linguistic Structure}\\
Verbosity (Response-level) & 54.57 & 234.61 & 106.37 & *** & 246.58 & 119.30 & *** & 323.80 & 145.48 & *** & 12844.50 & ***\\
Verbosity (Sentence-level) & 14.70 & 17.89 & 16.70 & *** & 18.15 & 18.10 & *** & 18.49 & 19.80 & *** & 5288.70 & ***\\
Readability & 2.33 & 7.94 & 2068.82 & *** & 7.20 & 1585.87 & *** & 6.52 & 1158.87 & *** & 2441.09 & ***\\
Repeatability & 0.10 & 0.16 & 26.27 & *** & 0.16 & 26.73 & *** & 0.15 & 23.39 & *** & 1415.12 & ***\\
Complexity & 9.47 & 15.35 & 28.75 & *** & 15.78 & 30.89 & *** & 15.78 & 30.84 & *** & 3303.52 & ***\\
\rowcollight \multicolumn{13}{l}{Linguistic Style}\\
Categorical Dynamic Index (CDI) & 2.52 & 7.22 & 10.00 & *** & 7.31 & 10.20 & *** & 7.87 & 11.39 & *** & 416.30 & ***\\
Formality & 0.57 & 0.95 & 39.88 & *** & 0.94 & 39.14 & *** & 0.92 & 37.5 & *** & 2279.6 & ***\\
Empathy & 0.70 & 0.65 & -913.11 & *** & 0.69 & -107.75 & *** & 0.70 & -11.24 & *** & 1172.37 & *** \\
\rowcollight \multicolumn{13}{l}{Adaptability to Query}\\
Semantic Similarity & 0.39 & 0.62 & 45.31 & *** & 0.62 & 45.24 & *** & 0.63 & 45.86 & *** & 3570.28 & ***\\
Linguistic Style Accommodation & 0.89 & 0.97 & 18.03 & *** & 0.97 & 18.08 & *** & 0.97 & 18.24 & *** & 2165.4 & ***\\
Diversity & 0.46 & 0.16 & -76.42 & *** & 0.15 & -78.91 & *** & 0.16 & -77.43 & *** & 4174.86 & ***\\
\bottomrule
\end{tabular}
\label{ai_compare}
\end{table}



\subsubsection{Linguistic Structure} For linguistic structure, we operationalized several measures of verbosity, readability, repeatability, and complexity that we describe below:

\para{Verbosity} serves a measure of detail and elaboration in communication, which is often associated with the effectiveness of support~\cite{saha2020causal,glass1992quality}.
We operationalized two types of verbosity at---1) \textit{response-level}: the total number of words per response, and 2) \textit{sentence-level}: the average number of words per sentence.
On average, AI responses were approximately 4 to 5 times longer and contained more words per sentence than \oc{} (mean=54.57) responses with statistical significance. 
A notable trend across the three AI configurations is the marked increase in verbosity with increased contextual information. 

\para{Readability} reflects how easily a reader can comprehend a given text. In health and online health contexts, it plays a crucial role in both expression and interpretation\cite{ernala2017linguistic, mcinnes2011readability}. 
We obtained the Coleman-Liau Index \( CLI \)~\cite{coleman1975computer} to assess readability, which evaluates character and word structure within a sentence. The \( CLI \) is calculated as 
\( CLI = (0.0588L - 0.296S - 15.8) \)
, where \( L \) represents the average number of letters per 100 words, and \( S \) represents the average number of sentences per 100 words. 
We found that AI responses show higher readability than \oc{} responses. 
Although higher readability indicates improved writing quality, it can also imply a greater educational requirement for comprehension.
Interestingly, adding more contextual information lead to a decrease in readability scores, with AI-3 showing the lowest readability among the AI responses. 
This could be indicative of the fact that adding additional context led responses to be marginally closer to human-written responses.

\para{Repeatability and Complexity} are syntactic measures that 
are associated with cognitive processes such as planning, execution, and memory~\cite{ernala2017linguistic}. 
Repeatability refers to the frequency of word reuse, where higher values may indicate lower communication quality due to redundancy. 
On the other hand, complexity, measured by the average length of words per sentence, influences how effectively ideas are conveyed, with greater complexity often associated with nuanced, precise, and detailed communication~\cite{kolden2011congruence}.
Again, \ai{} responses showed a significantly higher repeatability and complexity than \oc{} responses. 
That said, both of these measures remain largely similar across AI-1, AI-2, and AI-3 responses.

\subsubsection{Linguistic Style} Next, within linguistic style, we operationalized and compared across categorical-dynamic index (CDI), formality, empathy, and hopefulness in responses. 

\para{Categorical Dynamic Index (CDI)} is a bipolar linguistic measure that assesses writing style on a spectrum from categorical to dynamic~\cite{pennebaker2001linguistic}. 
We calculated CDI of each response by obtaining the parts of speech occurrences as per LIWC~\cite{tausczik2010psychological}.
A higher CDI value reflects a categorical writing style characterized by structured, abstract, and analytical expression, whereas a lower CDI signifies a dynamic or narrative style, emphasizing storytelling and fluidity. 
We found that \ai{} responses exhibited about 200\% higher CDI than \oc{} responses. This indicates that \oc{} members use a narrative and dynamic style of writing in responding to posts, whereas the AI uses a more categorical and analytical style of writing. 

\para{Formality} is a key sociolinguistic feature 
that reflects the level of sophistication, politeness, and relevance to linguistic norms~\cite{larsson2020syntactic}. Formal language is typically structured, grammatically precise, and commonly used in professional, academic, and official settings, whereas informal language adopts a more relaxed tone, often incorporating slang, colloquialisms, and abbreviations. 
To assess formality, we leveraged a RoBERTa-based formality classification model~\cite{babakov2023don,liu2019roberta} trained on Grammarly’s Yahoo Answers Formality Corpus (GYAFC)~\cite{rao2018dear}, achieving an ROC-AUC of 0.98 on benchmark datasets.
We found that formality is exhibited with a much higher extreme in AI responses (mean>0.92) than \oc{} responses (mean=0.57) with a statistically significant difference.
Increasing contextual information for the AI resulted in similar formality scores, with a slight decrease.


\para{Empathy} is a complex cognitive ability that allows individuals to understand and share the emotions and perspectives of others, playing a crucial role in supportive communication by fostering emotional connection and validation~\cite{herlin2016dimensions}. We employed a RoBERTa-based model trained on empathetic reactions to news stories~\cite{buechel2018modeling}.
Interestingly, \ai{} responses showed a lower empathy than \oc{} responses. 
However, with the addition of more contextual information, empathy scores get higher in \ai{} responses, getting closer to \oc{} responses.

\subsubsection{Adaptability to Query} Finally, we operationalized measures in how the responses adapted to the queries in terms of semantic similarity, linguistic style accommodation, and diversity. 

\para{Semantic Similarity} measures the extent to which a response is topically and contextually similar to a post. We computed the cosine similarity between the 384-dimensional embeddings of posts and responses using a transformer-based language model, \textit{all-MiniLM}~\cite{wang2020minilm}.
We found that AI responses showed a significantly higher semantic similarity than \oc{} responses. 
We also noted a marginal increase in semantic similarity with added context in prompting the AI.

\para{Linguistic Style Accommodation} goes beyond content similarity and evaluates how well a response stylistically matches its query, focusing on non-content words such as function words and pronouns~\cite{danescu2011mark}.
Prior research showed that adapting to a user's writing style can improve online support~\cite{saha2020causal, sharma2018mental}.
We computed the occurrences of these parts of speech using the LIWC~\cite{tausczik2010psychological}. Then, we obtained the vector representations of posts and corresponding responses on the occurrences of these parts of speech and measured the cosine similarities to quantify the linguistic style accommodation.
We found that AI responses show higher linguistic style accommodation than \oc{} responses, and almost perfectly match the linguistic style of the queries (mean=0.97). 

\para{Diversity} or creativity refers to the uniqueness and variation in responses, and greater diversity is known to be associated with greater effectiveness in psychotherapy and social support~\cite{althoff2016large, norcross2018psychotherapy}.
To measure diversity, we computed centroid vectors from word embeddings in a 384-dimensional space using the \textit{all-MiniLM model}~\cite{wang2020minilm} and assessed the cosine distance of individual responses from these centroids. 
A greater distance indicated higher linguistic diversity, reflecting more varied and creative responses. 
We found AI responses show a much lower diversity than \oc{} responses. This might be indicative of the aspect that AI tends to reuse and repurpose similar suggestions across several responses. 
Therefore, even though AI can generate coherent and contextually relevant responses, the responses lack diversity. On the other hand, online community members are likely to provide experience-based suggestions and personal narratives, exhibiting higher diversity.

%% file: 7Aim4.tex
\subsection{Expert Evaluation of AI Responses to \sci{}: Anticipating Concerns and Harms.}

It is critical to examine AI's potential benefits and limitations in responding to individuals experiencing \sci{}. 
We qualitatively explored the nuances in the AI responses and identified whether these responses could lead to possible harm. 
We obtained a random sample of 200 posts (and corresponding AI responses) and had these expert-appraised by our psychologist co-authors to provide detailed assessments.



The evaluation of these AI-generated responses was conducted based on the principles of Cognitive Behavioral Therapy (CBT), with a particular focus on identifying whether the responses unintentionally reinforced cognitive distortions such as catastrophizing and overgeneralization. 
This approach aligned with existing research on the role of cognitive distortions in \sci{}~\cite{buscher2020internet}. Additionally, the tone and structure of the responses were also assessed to determine their alignment with supportive and constructive communication strategies. This evaluation was guided by established suicide prevention frameworks, including the SAFE-T model (SAMHSA, 2009)~\cite{safet}, which emphasizes structured, empathetic engagement to enhance safety and directly address suicidal thoughts. Furthermore, the AI responses were compared with established best practices from the literature on responding to \sci{}, particularly within the context of online communities. In particular, the psychologist co-authors applied the Collaborative Assessment and Management of Suicidality (CAMS) model~\cite{comtois2023reducing}, emphasizing collaboration and individualized care, while evaluating the AI's ability to reflect similar collaborative principles.
Based on their comments, we grouped the observations into the following key themes:

\para{Emotional Alignment and Response Effectiveness:}
AI-3, which was prompted with the \sci{} post, \sci{} category, and linguistic characteristics of supportive responses, exhibited responses with stronger emotional alignment with user distress compared to AI-1 and AI-2. 
It employed more explicitly empathetic language, such as ``I truly feel for you'' or ``I can imagine how difficult this must be.'' These linguistic markers suggest an effort to build rapport and validate emotions, which can be crucial in fostering trust in digital interventions. However, while AI-3’s responses were perceived as more compassionate, the improvements over AI-1 and AI-2 were often subtle rather than substantial. Despite its increased emotional alignment, AI-3, like its counterparts, sometimes defaulted to generalized supportive statements such as ``I'm sorry to hear that,'' which could feel impersonal. In a small proportion of cases, AI-generated responses contained no additional text beyond a few such broadly supportive statements. The chatbot’s tendency to rely on pre-formulated expressions limited its ability to engage meaningfully in nuanced conversations, highlighting a fundamental challenge in AI-mediated crisis support---balancing emotional alignment of responses with conversational depth.

\para{Personalization and Trust-Building:}
A recurring concern across all AI models was the lack of personalization in responses. While none of the AI-generated replies were overtly harmful, many lacked specificity in addressing the unique concerns of each user. This was particularly evident when AI-generated responses failed to acknowledge key details in posts, such as prior negative experiences with mental health professionals or distrust of medical systems. For instance, when users expressed disillusionment with therapy, the chatbot frequently recommended seeking professional help without adapting its response to account for the user’s reservations. 


\para{Shifting Between Supportive Listening and Intervention.}
One notable finding was the variation in how AI models determined when to shift from empathetic engagement to recommending crisis resources. Posts containing explicit references to prior suicide attempts or methods did not always elicit a shift in AI responses toward immediate intervention. While AI-2 and AI-3 were more likely to encourage users to trust hospital-based treatment providers, AI-1 occasionally prompted further conversation without recommending professional support. This inconsistency raises ethical considerations regarding AI-driven risk assessment. Future research should explore optimal strategies for balancing empathetic engagement with timely intervention while avoiding responses that feel formulaic or dismissive.

\para{Validation Without Reinforcing Harmful Cognitions.}
An encouraging finding was that none of the AI models explicitly reinforced \sci{} or validated harmful cognitive distortions. 
However, subtle differences emerged in how the models addressed suicidal thoughts. For instance, in response to posts asserting that suicidal thinking is ``normal,'' AI-2 explicitly countered this notion by stating, ``Feeling suicidal is \textit{not normal}, and wanting to harm yourself isn't something you should cope with in silence.'' 
In contrast, AI-3 was less direct in challenging such assertions. While pushing back against harmful beliefs can be beneficial, the way this is done matters. Responses that feel overly clinical or detached---such as AI-2's phrasing in some cases---may risk alienating users who seek emotional validation. 
Future iterations of AI-driven crisis support should focus on balancing between validating distress and gently guiding users toward reframing harmful thoughts in a non-confrontational manner.

\para{Variability Across AI Models.}
Although AI-3 displayed a slight tendency toward more expressive empathy, the overall differences between the three models were not always stark. The variation in responses was often more attributable to the nature of the user's post rather than fundamental differences in AI architecture. Given this variability, further analysis is required to determine whether specific fine-tuning strategies consistently enhance AI-driven support systems.

\para{Potentially Less Effective Responses:}
Despite their generally supportive nature, some AI responses were less helpful due to issues in phrasing or tone. Responses such as ``Our minds can sometimes trick us into believing things that aren't true, especially when we're feeling down'' (AI-1) risked sounding dismissive rather than reassuring. Similarly, AI-generated responses about the user in the third person (e.g., ``You mentioned that you aren't scared after your attempt. This can be deeply concerning, as it might indicate an increased risk of attempting again.'') could feel impersonal and detached.

\para{Balancing Response Length and Engagement:}
A final consideration is the optimal length and depth of AI-generated responses. While generic responses were sometimes perceived as less helpful, overly lengthy replies also risked being impractical. Users who posted brief messages often received disproportionate length responses, potentially making engagement feel unnatural. A more conversational approach—where response length aligns with the user's post and includes follow-up questions—may enhance interaction quality while preserving the chatbot's role as a supportive entity rather than an information dispenser.

%% file: 8discussion.tex
\section{Discussion}

\subsection{Theoretical Implications}


\subsubsection{Applicability of \its{} in online disclosures of suicidal ideation}
This study underscores the relevance and adaptability of the Interpersonal Theory of Suicide (\its{}) in online settings. 
We found a high overlap between manual and computational \its{} labels in our dataset
Our findings indicated that posts classified under \its{} categories exhibit cognitive and emotional patterns consistent with those described in \its{}~\cite{chu2017interpersonal, van2010interpersonal, joiner2010interpersonal}, reinforcing its applicability in analyzing large-scale data in online settings.
\autoref{theme_distribution} further reinforced that posts classified as lethally suicidal had higher weaknesses and pain, planning and attempts, self-harm, and methods and tools. 
This emphasizes how individuals at high risk for suicide often express intense emotional distress through their online interactions.
These findings align with prior theoretical understanding that suicide attempters likely have higher pain tolerance and a history of prior attempts compared to those with ideation alone~\cite{posner2008columbia, juhnke2007path, greaves2024linguistic}.

\subsubsection{Alignment with theoretical frameworks of suicidal ideation}
Along the lines of the above, our linguistic analyses in \sci{} disclosures helped us identify the major concerns and topics.
Our observations not only support \its{}, but also converge with other validated theoretical frameworks explaining the progression of \sci from ideation to action, such as the Three-Step Theory (3ST)~\cite{klonsky2021three} and Integrated Motivational-Volitional Model (IMV)~\cite{o2018integrated}.
The 3ST traces suicide risk through distress, disconnection, and acquired capability, reflected in posts on psychological pain and methods. 
Likewise, the IMV model highlights the voluntary factors that differentiate ideators from attempters, such as prior attempts and increased pain tolerance---as also supported by our findings.

\subsubsection{Understanding responses to \sci{}}
Our work offers an empirical analysis of supportive responses to various types of \sci{}.
In particular, in Aim 2, we analyzed the language of the responses to different types of \sci{} labeled posts, revealing distinct patterns as a function of \its{} constructs.
Notably, the majority of responses exhibited traits of care, empathy, and advice-giving, as shown in~\autoref{table:sage_lethal}. Common phrases such as ``that's okay'', ``life problems'', ``change lot'', and ``friend like'' frequently appeared in responses to both lethal and non-lethal \sci{} posts. 
This suggests that while empathy is a primary motivator in these responses, there may be a tendency for responders to generalize or downplay the severity of the distress conveyed in these posts. 
These observations align with prior research on highlighting the role of supportive language in mitigating distress~\cite{fu2013responses, bjarehed2023different, o2018rate}.
Further, responses to perceived burdensomeness often contained a higher degree of positive affect and references to friendships, suggesting that respondents employ affirmative language to provide reassurance and mitigate concerns about interpersonal relationships. 
In contrast, responses to acquired capability frequently referenced themes of religion, achievement, and health, reflecting existential contemplation and considerations of physical resilience. 
These findings advance our knowledge that different types of \sci{} disclosures may trigger specific emotional responses, which can provide important insights into the diverse needs of individuals in crisis. 


\subsubsection{How Unique is Suicidal Ideation in Online Spaces Compared to Traditional Contexts?}
The theoretical framing of \sci{} is rooted in traditional (offline) contexts, shaping our understanding of risk factors and behaviors. Our research offers overlapping as well as complementary insights into the distinct dynamics of online suicidal ideation, particularly how perceptions of belongingness and burdensomeness---key constructs in \its{}---differ in online discourse.
Unlike offline settings, online spaces provide anonymity, asynchronous interaction, and broader peer validation, which may influence how individuals express and receive support around suicidal thoughts. For example, the ability to engage in persistent, text-based interactions allows for reflections that might not be shared in face-to-face interactions, and users may construct narratives around suicidal ideation in ways that shape their sense of social connectedness differently. Additionally, AI chatbots introduce a non-human yet responsive conversational dynamic, which could challenge or reinforce \its{} assumptions about interpersonal support and validation in suicide risk.
Overall, by applying \its{} to online contexts, our study advances the theoretical knowledge by considering how digital affordances alter the experience of perceived burdensomeness and thwarted belongingness. 

\subsubsection{Limitations of \ai{} for \sci{} support}
Our study conducted a preliminary investigation into \ai{} support for \sci{}. 
We found that AI responses are structured and articulate, but they may lack emotional depth, diversity, and connectivity on a human level, as also evaluated by our psychologist co-authors. 
The tone of these AI chatbots may not always align with user expectations, as some users might prefer a conversational and empathetic style~\cite{sakirin2023user}.
Therefore, a key limitation of AI chatbots is their inability to share personal experiences, affecting the perceived authenticity of their responses~\cite{rostami2023artificial,chandra2024lived}. Despite demonstrating computational empathy, AI lacks the lived experience that fosters trust in peer support. Users seeking structured, factual responses may find AI effective, while those desiring personal connections may perceive its support as inadequate~\cite{li2024developing}. 
This highlights the need for AI systems to evolve beyond simple empathy and towards a deeper, context-aware understanding of users' emotional states. 
These findings thereby suggest the need for further research into balancing AI's structured approach with emotionally supportive engagement while maintaining transparency.



\subsection{Practical and Design Implications}

\subsubsection{Designing online mental health spaces by adopting \its{}}
Online mental health communities, such as on Reddit, can enhance their support systems and foster stronger connections by incorporating the \its{} framework into the design of their platforms~\cite{wang2023metrics}. By recognizing perceived burdensomeness and thwarted belongingness as critical risk factors for suicidal ideation, \its{} offers a foundational approach for creating effective community engagement strategies. Platform design can integrate features such as tailored response templates and alert tools to detect when users express these feelings, enabling moderators and peer supporters to offer validation, reassurance, and support. 
Further, community guidelines and platform interactions can be structured to reduce isolation and feelings of burdensomeness, fostering a more inclusive and supportive environment. Additionally, the design of dedicated discussion threads or mentorship programs can facilitate sustained engagement to combat loneliness. Training moderators to identify posts with high \its{}-related risk factors, alongside trigger alerts and real-time interventions, allows platforms to act quickly when acute distress is detected. Integrating \its{} principles into the design of online platforms helps cultivate a sense of belonging, offering at-risk individuals better support and a stronger sense of community.


\subsubsection{AI's potential in timely \sci{} interventions} 
While AI-generated responses demonstrated linguistic sophistication and adaptability, their lack of lived experience limits emotional engagement in highly sensitive contexts. 
AI responses maintain a structured and professional tone, which may feel impersonal in emotionally charged conversations. 
However, AI can serve as a first responder in mental health apps by de-escalating distressing situations and providing structured insights for clinicians. 
AI-generated summaries of user interactions can also assist professionals in tailoring interventions, identifying crisis patterns, and prioritizing cases based on urgency, effectively combining AI’s analytical capabilities with human empathy.

Further, AI chatbots can play a vital role in crisis stabilization, particularly when immediate human support is unavailable. 
Designing AI systems with adaptive response variability---where phrasing and content evolve based on prior user interactions---can enhance engagement while maintaining reliability. 
Additionally, incorporating contextual memory can allow AI to track previous interactions and avoid redundant or overly generic responses.
A user-centric approach to designing AI systems that dynamically adjust based on perceived distress levels could improve their effectiveness in such sensitive interactions.


\subsubsection{Training mental health volunteers}
This work bears implications for designing training approaches for mental health volunteers, particularly in light of the growing need for skilled support. 
For instance, AI chatbots can serve as an effective tool for training mental health volunteers by providing real-time, scenario-based simulations. By analyzing patterns in high-quality and supportive responses from online support communities, AI can generate realistic conversational scenarios that help volunteers develop essential communication skills. These chatbots can offer interactive role-playing exercises, guiding trainees on how to craft supportive, empathetic, and contextually appropriate responses to individuals experiencing \sci{}. By integrating linguistic and psychological insights, such systems can enhance volunteers' ability to recognize distress signals and tailor their responses accordingly.

\subsection{Ethical Implications}\label{sec:ethical_implications}

\subsubsection{Risk of Misinterpretation and Harm with Automated Approaches}  
Our work on suicidal ideation presents sensitive ethical, methodological, and epistemological challenges, particularly regarding individuals at risk. Automating the identification of suicidal ideation in online content raises concerns about misinterpreting distress or reinforcing harmful stigmas, especially when individuals may not want their struggles exposed or quantified. We recognize the ethical tension where individuals experiencing suicidal thoughts may seek help but do not want their distress to be labeled or identified. Furthermore, algorithmic methods could unintentionally overlook individuals in need (false negatives) or misclassify their emotional states, resulting in inadequate support~\cite{kaur2022didn,kawakami2023wellbeing,roemmich2021data}.
Additionally, using automated systems to allocate resources based on detected suicidal ideation poses challenges. Our approach, based on publicly shared text, might fail to identify individuals who are in critical need of support or misjudge the severity of their crisis. Suicidal ideation manifests differently across individuals, making it difficult for an algorithm to allocate resources effectively or appropriately. These concerns emphasize the importance of using AI ethically to avoid harm and ensure that vulnerable individuals receive timely, personalized assistance.

These concerns are further compounded by the potential misuse of such data and computational approaches by external parties like insurance companies or targeted advertisers. 
Publicly available online data could be exploited for commercial gain, with companies potentially using insights about mental health crises to deny insurance coverage or target vulnerable individuals with exploitative advertising. 
This underscores the need for careful ethical considerations to prevent harm and ensure that vulnerable individuals receive timely, personalized assistance while safeguarding their privacy and autonomy.

\subsubsection{Algorithmic Bias and Equity in AI Support.}
A critical concern in deploying AI to respond to individuals experiencing suicidal thoughts is the risk of misinterpretation and the potential for harm. LLMs are known to exhibit hallucinations---instances where the model generates factually incorrect, misleading, or fabricated information. In the context of suicide prevention, such inaccuracies can have severe consequences, particularly if an AI misinterprets the individual's distress, provides inappropriate responses, or offers guidance that exacerbates their crisis~\cite{park2024toward}. Ethically, this raises significant concerns about accuracy, as misinformed guidance can lead to further emotional distress or even encourage harmful behaviors.

While our study does not explore the technicalities of AI biases in-depth, but opens up ethical concerns in this regard. 
AI models, trained on broad datasets not specifically designed for mental health, may carry biases that reflect societal inequalities and cultural disparities. 
The widespread availability of AI chatbots, such as ChatGPT, Gemini, and others, allows individuals to ask questions about their mental health without clear ethical guardrails. This raises concerns about the quality and safety of the information provided, as users may inadvertently receive misleading or harmful advice without adequate oversight or professional support.
This can also be particularly harmful for marginalized communities~\cite{zhou2024bias}. For example, an AI trained on Western English data may accurately understand linguistic and cultural variations in non-Western or non-English-speaking populations~\cite{kamruzzaman2023investigating}. Such biases can result in misjudgments of risk, emphasizing the need for further studies as well as policies to prevent and mitigate the potential harms of AI in the context of digital mental health interventions.


\subsubsection{Emotional Authenticity and Transparency.}
Ensuring that AI models seem authentic while still preserving transparency is crucial for fostering trust and ethical engagement with their users. Individuals seeking support for \sci{} often require responses that demonstrate genuine empathy, understanding, and emotional depth~\cite{yang2022influence,de2017language,saha2020causal}. While AI models can mimic human conversational traits like empathy, and emotional connection, we observed from this study that it does not correlate to authenticity as explained by the psychologists (e.g., providing responses that are supportive in tone, but not fully acknowledging or addressing key details specific to the post). 
However, it is essential to ensure that AI not only mimics human-like interaction but also maintains transparency between the user and the system. 
Users may perceive AI-generated responses as human-like, leading to misplaced trust and potential harm. Therefore, transparency is crucial to managing these risks. 
A notable example is KOKO, an AI-powered mental health chatbot~\cite{koko}---which sparked controversy by not disclosing to its users that it provided AI-generated mental health support. Upon discovery, users felt deceived, leading to an ethical backlash~\cite{koko}. This highlights the importance of fostering system transparency to ensure informed engagement with technology and maintaining ethical standards in mental health support.

\subsubsection{Privacy, Data Sensitivity, and User Consent.} 
Deploying AI chatbots for suicidal ideation raises concerns about privacy and data security. Given the deeply personal and distressing nature of \sci{}, individuals seeking support must be assured that their data is handled with the utmost confidentiality and security. 
Mishandling such data can lead to breaches, exposure, or legal consequences. 
These technologies need to adopt techniques such as differential privacy, encryption, and anonymization to protect user identities, while adherence to regulations like General Data Protection Regulation (GDPR)~\cite{voigt2017eu} and Health Insurance Portability and Accountability Act (HIPAA)~\cite{moore2019review} is essential in such sensitive mental health contexts. 
These technologies need to adopt a proper consent process, and also ensure that users are well-informed about data usage, retention, and third-party access, with the option to opt-out or delete their data.


\subsection{Limitations and Future Directions}


Our study has limitations, which also suggest interesting future directions. 
In the case of suicidal ideation, our study is limited by what can be observed from online data alone. 
Despite corroborating our findings with psychologists specializing in suicidal ideation, the lack of complementary information---such as clinical assessments or physiological data---prevents us from making clinical claims about \sci{}.
Our data lacks formal clinical validation based on established diagnostic frameworks such as the DSM-5~\cite{dsmfive2013} or RDoC~\cite{cuthbert2013toward}. 
While our findings offer valuable insights, we caution against drawing direct clinical or diagnostic inferences. 
Nonetheless, this work can serve as a foundation for future research, including replication studies in clinical settings.

Further, our work does not empirically assess the effectiveness of AI-generated responses on individuals experiencing \sci{}. 
As this remains a nascent field with many unknowns, our study utilized retrospectively collected data, combining quantitative analyses with expert-led qualitative assessments of relevance and supportiveness in the language of responses.
Our work inspires future research in understanding the effectiveness of AI in \sci{} intervention through deployment and experimental studies (with sufficient human supervision) in terms of how these technologies can impact the wellbeing of individuals.
Additionally, future research can incorporate direct feedback from various stakeholders, including mental health professionals, individuals experiencing \sci{}, moderators, and platform owners, in designing effective online support interventions.

Our study likely suffers from self-selection bias, only studying the data of individuals who choose to disclose \sci{} on online communities, and may not encapsulate the entire spectrum of individuals undergoing \sci{} and may need to be helped with digital mental health interventions.
Again, since the linguistic features of suicidal thought disclosures---and responses to those disclosures---likely vary across online communities beyond \swatch{} and across other platforms, future research can investigate diverse online interactions, including private messaging platforms (e.g., WhatsApp) and online therapy settings.

%% file: 9conclusion.tex
\section{Conclusion}

This paper adopted the Interpersonal Theory of Suicide (\its{}) as an analytical lens to examine suicidal ideation (\sci{}) in online spaces. 
We developed a computational framework based on distant supervision based learning and iterative-codebook based similarity matching to identify \sdim{}s and \srsk{}s of \its{} in 59,607 posts from \swatch{} on Reddit.
Then, we conducted topic modeling to understand what is discussed in these \sdim{}s and \srsk{}s. 
We found that posts exhibiting high-risk of suicidal intent expressed planning, attempts, and pain tolerance. 
Next, we examined the responses to different types of \sci{} through psycholinguistic and content analyses. 
Our analyses revealed that responses to \sci{} posts consistently conveyed empathy and support. 
However, responses to Thwarted Belongingness showed more negativity and uncertainty, whereas those to Perceived Burdensomeness posts prompted hesitant positivity, and those to acquired capability posts elicited urgency and sharing personal experiences.
Finally, we examined AI chatbots' role in providing supportive responses, finding that while they enhanced readability, formality, and coherence, they also exhibited high verbosity, complexity, and repetition, lacking emotional depth and personalization.
We discussed the implications of this work in adopting a theory-driven lens in understanding suicidal ideation, and toward ethical integration of AI in complementing, but not replacing human expertise in digital mental health interventions.